\documentclass[12pt]{article}

\usepackage{amsmath,graphicx}
\usepackage{multirow}
\usepackage{amsfonts}
\usepackage{amssymb}
\usepackage{amscd}
\usepackage{cite}
\usepackage{amsmath}
\usepackage{bbm}
\usepackage{amscd,bbold}

\def\hybrid{\topmargin -20pt    \oddsidemargin 0pt
        \headheight 0pt \headsep 0pt
        \textwidth 6.25in       
        \textheight 9.5in       
        \marginparwidth .875in
        \parskip 5pt plus 1pt   \jot = 1.5ex}

\hybrid

\numberwithin{equation}{section}
\numberwithin{table}{section}

\renewcommand{\Re}{\operatorname{Re}}
\renewcommand{\Im}{\operatorname{Im}}
\newcommand\e{\mathrm{e}}
\newcommand\iu{\operatorname{i}}
\newcommand\diff{\mathrm{d}}
\newcommand{\vol}{{\operatorname{vol}}}
\newcommand{\tr}{{\operatorname{tr}}}
\newcommand{\otimesF}{{\,\odot\,}}
\newcommand{\otimesA}{{\,\circ\,}}

\newcommand{\bbR}{\mathbb{R}}

\def\rep#1{\mbox{{\bf #1}}}

\DeclareMathOperator{\SL}{\mathit{SL}}

\DeclareMathOperator{\E7}{\mathit{E}_{7}}
\DeclareMathOperator{\Es7}{\mathit{E}_{7(7)}}

\newcommand{\Tsub}{ O(6,6) \times \SL(2,\bbR)}

\DeclareMathOperator{\Ex6}{\mathit{E}_{6(2)}}
\newcommand{\SLE}{\SL(8,\bbR)}

\newcommand{\be}{\begin{equation}}
\newcommand{\ee}{\end{equation}}
\newcommand{\beq}{\begin{equation}\begin{aligned}}
\newcommand{\eeq}{\end{aligned}\end{equation}}
\newcommand{\bi}{\begin{itemize}}
\newcommand{\ei}{\end{itemize}}
\newcommand{\bea}{\begin{eqnarray}}
\newcommand{\eea}{\end{eqnarray}}
\newcommand{\ba}{\begin{array}}
\newcommand{\ea}{\end{array}}
\newcommand{\bt}{\begin{tabular}}
\newcommand{\et}{\end{tabular}}
\newcommand{\bc}{\begin{center}}
\newcommand{\ec}{\end{center}}

\newcommand{\hm}{\hat m}
\newcommand{\hn}{\hat n}
\newcommand{\nn}{\nonumber}
\newcommand{\TsubO}{O(6,6) \times \SL(2,\bbR)_{O}}
\newcommand{\TsubOp}{O(6,6) \times \SL(2,\bbR)_{O(p+3)}}

\begin{document}

\begin{titlepage}

\begin{center}

\rightline{\small IPhT-t12/099}

\vskip 2cm

{\Large \bf Generalized $N=1$ and $N=2$ structures} \\
\vskip 0.5cm
{\Large \bf in M-theory and type II orientifolds  }

\vskip 1.5cm

{\bf Mariana Gra\~na and Hagen Triendl}\\

\vskip 0.5cm

{\em Institut de Physique Th\'eorique, CEA Saclay,\\
Orme de Merisiers, F-91191 Gif-sur-Yvette, France}

\vskip 0.5cm

{\tt mariana.grana@cea.fr, hagen.triendl@cea.fr} \\

\end{center}

\vskip 2cm

\begin{center} {\bf ABSTRACT } \end{center}

We consider M-theory and type IIA reductions to four dimensions with $N=2$ and $N=1$ supersymmetry and discuss their interconnection. Our work is based on the framework of Exceptional Generalized Geometry (EGG), which extends the tangent bundle to include all symmetries in M-theory and type II string theory, covariantizing the local U-duality group $\Es7$.
We describe general $N=1$ and $N=2$ reductions in terms of $SU(7)$ and $SU(6)$ structures on this bundle and thereby derive the effective four-dimensional $N=1$ and $N=2$ couplings, in particular we compute the K\"ahler and hyper-K\"ahler potentials as well as the triplet of Killing prepotentials (or the superpotential in the $N=1$ case).
These structures and couplings can be described in terms of forms on an eight-dimensional tangent space where $SL(8) \subset E_7$ acts, which might indicate a description in terms of an eight-dimensional internal space, similar to F-theory.
We finally discuss an orbifold action in M-theory and its reduction to O6 orientifolds, and show how the projection on the $N=2$ structures selects the $N=1$ ones. We briefly comment on new orientifold projections, U-dual to the standard ones.

\vfill

\today

\end{titlepage}


\tableofcontents
\newpage

\section{Introduction}\label{section:Intro}
It has been a long-standing problem to understand more general flux backgrounds in string theory. The most successful tool for this task has been supersymmetry, since it reduces the second-order equations of motion to first-order supersymmetry equations that are much simpler to solve and provides non-renormalization theorems for certain sets of couplings. The simplest string backgrounds are Ricci-flat geometries that admit a number of covariantly constant spinors. However, in the presence of fluxes, the first-order supersymmetry equations become more involved, as all bosonic fields of string theory appear in these equations. The global bosonic symmetries of the massless string spectrum then can help to write these first-order equations in a simpler form. This is the approach of generalized geometry. The introduction of a generalized tangent bundle covariantizes these symmetries by combining the diffeomorphisms and gauge transformations of the gauge fields into a single vector bundle. The metric on this bundle then includes, besides the metric, the other massless bosonic fields. The idea was first applied to the T-duality group that transforms the metric and the B-field into each other. This description goes by the name of Generalized Complex Geometry \cite{Hitchin}.

More recently, the formalism was extended to the U-duality group of type II string theory to also include its Ramond-Ramond fields and to M-theory, covariantizing its three-form gauge field \cite{Hull,Pacheco:2008ps,Grana:2009im,Coimbra:2011ky}. In both cases the covariantized symmetry group relevant to compactify to four dimensions is $\Es7$, and the generalized tangent bundle of type IIA is recovered from the one of M-theory by dimensional reduction. All possible supersymmetric backgrounds of type II and M-theory can then be understood as (generalized) $G$-structure backgrounds, where $G$ is, depending on the number of supercharges, a subgroup of $SU(8)$ (the maximal compact subgroup of $\Es7$).\footnote{Often these are called generalized $G$-structures as they are structures on the generalized tangent space, as opposed to usual $G$-structures on the tangent space.} The first-order supersymmetry equations can then be written in a covariant way \cite{Grana:2011nb,Grana:2012ea,Waldram:2012}. Furthermore, $G$-structure backgrounds that solve the equations of motion but not the first-order supersymmetry conditions can usually be understood as backgrounds with spontaneously broken supersymmetry. In this sense, generalized geometry serves as an off-shell supersymmetric formulation for general $G$-structure backgrounds.

Generalized $G$ structures are characterized by fundamental objects (defined by bi-spinors) in certain representations of $\Es7$ such that their common stabilizer is exactly $G$. These fundamental objects determine the generalized metric (which is stabilized by $SU(8)$) and thereby the entire bosonic supergravity background. The first-order supersymmetry conditions are resembled by first-order differential equations on them. Moreover, the couplings of the off-shell supersymmetric formulations are given in terms of tensorial combinations of these fundamental objects into $\Es7$ singlets.

A major aim of this work is to understand $SU(7)$ and $SU(6)$ structures in M-theory and type II, as they correspond to backgrounds with the off-shell structure of four-dimensional $N=1$ and $N=2$ supergravity. $SU(7)$ structures in M-theory have already been discussed to some extend in \cite{Pacheco:2008ps}, $SU(6)$ structures in type IIA have already been discussed in \cite{Grana:2009im}.
Of particular interest in our discussion will be the existence of an eight-dimensional intermediate bundle between the tangent and the generalized tangent bundle that admits the action of $SL(8)\subset \Es7$. This might indicate the existence of a twelve-dimensional theory (similar to F-theory) in which some of the charges will be geometrized. $SU(7)$ and $SU(6)$ structures are related to $Spin(7)$ and $SU(4)$ structures in $SL(8)$.

After we briefly review the $\Es7$-covariant formalism of exceptional generalized geometry in Section~\ref{sec:EGG}, we will discuss $SU(7)$ and $SU(6)$ structures in M-theory in Section \ref{sec:SUnstr}. In particular, we will discuss all classical couplings in these backgrounds. Subsequently, we will relate these results to the equivalent structures in type IIA string theory in Section~\ref{sec:IIA}. While $SU(6)$ structures descend in a straight-forward way, $SU(7)$-structures are related to O6-orientifold backgrounds in type IIA.
Moreover, $SU(6)$ and $SU(7)$ structures should be related by involutions that project out half of the supersymmetry, such as orbifolding and orientifolding in M-theory and type II string theory. In Section~\ref{sec:Orbifolding} we will show how an $SU(7)$ structure is obtained from an $SU(6)$ structure via such involutions. In particular, we will determine the $N=1$ couplings in terms of the $N=2$ parent theory. Finally, we will identify the involutions given by standard orbifolding and orientifolding in M-theory and type II string theory and discuss a few new involutions, for instance an involution creating objects with tension and charge opposite to those of NS5-branes. We conclude with a summary of the results and some outlook. Appendix \ref{App:E7} contains all the relevant formuli concerning $\Es7$ representations, and Appendix \ref{App:computations} presents the details of some of the calculations done along the paper.



\section{Exceptional Generalized Geometry (EGG)}
\label{sec:EGG}
In this section we review the basic concepts of Exceptional Generalized Geometry (EGG), emphasizing the role of the eight-dimensional intermediate tangent bundle $T_8$.
The idea of EGG is to covariantize the U-duality group $\Es7$ in M-theory and type II compactifications to four dimensions. Though only torus compactifications admit globally the action of the U-duality group, locally any background admits it, as the tangent space is isomorphic to $\mathbb R^7$ ($\mathbb R^6$). In EGG the internal seven- (or six-)dimensional tangent bundle of an M-theory (type II) compactification to four dimensions is enlarged to a 56-dimensional exceptional generalized tangent bundle such that the U-duality symmetry group $\Es7$ acts linearly on it. Thereby, the U-duality group promotes to a geometric action on this bundle. As the U-duality group maps all bosonic supergravity degrees of freedom into each other, EGG gathers them all in a metric on this exceptional generalized tangent bundle. The patching of the exceptional generalized tangent bundle \cite{Pacheco:2008ps} resembles the global aspects of the compactification.
More details can be found in \cite{Hull,Pacheco:2008ps,Grana:2009im,Coimbra:2011ky}.

\subsection{An eight-dimensional tangent space $T_8$}

In the case of compactifications of type II, the exceptional tangent bundle  combines the string internal momentum and winding charges (6+6 elements), their magnetic duals (another 6+6) as well as all the D-brane charges (32 elements). These together form the fundamental ${\bf 56}$ representation of $\E7$. In M-theory, it is a result of combining momentum and its dual (Kaluza-Klein monopole charge) (7+7) together with M2 and M5-brane charges (21+21). These charges can be combined into $SL(8, \mathbb R)$ representations. We can think of this group as acting on an 8-dimensional tangent bundle $T_8$, which will be split into $7+1$ for M-theory, and further split into $6+1+1$ for type IIA. Of course there is a priori no eight-dimensional manifold with a tangent bundle $T_8$ appearing in M-theory or type IIA. Therefore, $T_8$ should be seen as some kind of generalized tangent bundle. In terms of $SL(8, \mathbb R)$ representations, the fundamental of $\E7$ decomposes as
\begin{equation}\label{decomposition_sl8_fund} \begin{aligned}
 E =&\Lambda^2 T_8 \oplus \Lambda^2 T_8^* \ , \\
  {\bf 56} = & {\bf 28} \oplus {\bf 28'}  \ .
  \end{aligned}
\end{equation}
Similarly, for the adjoint we have
\begin{equation} \label{decomposition_sl8_ad} \begin{aligned}
 A = &(T_8 \otimes T_8^*)_0 \oplus \Lambda^4 T_8^* \ , \\
  {\bf 133} = & {\bf 63} \oplus {\bf 70}  \ ,
  \end{aligned}
\end{equation}
where the subscript $0$ denotes traceless.
We will also need the ${\bf 912}$  representation, which  splits according to
\begin{equation} \label{rep912}\begin{aligned}
 N = & S^2 T_8 \oplus (\Lambda^3 T_8 \otimes T_8^*)_0 \oplus S^2 T_8^* \oplus (\Lambda^3 T_8^* \otimes T_8)_0 \ , \\
 {\bf 912} = & {\bf 36} \oplus {\bf 420} \oplus {\bf 36'} \oplus {\bf 420'} \ ,
\end{aligned} \end{equation}
where $S^2$ denotes symmetric two-tensors.

When we later consider spinors it is also useful to use the maximal compact subgroups of the groups that are involved. The maximal compact subgroup of $\E7$ is $SU(8)$, and the group-theoretical decompositions are completely analogous to the $SL(8,\mathbb{R})$ case and are given by (\ref{decomposition_sl8_fund}-\ref{rep912}). Note though that the $SU(8)$ that transforms the spinors is not the compact subgroup $SU(8)$ of $SL(8)$ that acts on $T_8$. Nevertheless the two $SU(8)$ subgroups are related by some $E_{7(7)}$ transformation and the decomposition of $E_{7(7)}$ representations is the same in both cases. More details can be found in Appendix~\ref{sec:SU8grouptheory}.
When we then consider $SL(8,\mathbb{R})$, spinors transform under the corresponding spin group $Spin(8)$ and its maximal compact subgroup $SO(8)$. Note that for $Spin(8)$, we can impose a Majorana-Weyl condition on the spinor. The Weyl-spinors are in one-to-one correspondence to $Spin(7)$ spinors that are considered in the M-theory compactification.

\subsection{M-theory and $GL(7)$ decompositions}

For compactifications of M-theory on seven-dimensional manifolds, we should decompose further the $SL(8)$ into $GL(7)$ representations, or in other words split the 8-dimensional dual vector bundle $T_8^*$ into a 7-dimensional one $T_7^*$, plus a scalar piece. Choosing an overall power of the 7-dimensional volume form to get the correct embedding in $SL(8)$ (see more details in \cite{Pacheco:2008ps}), we get
\begin{equation} \label{decomposition_gl7_V}\begin{aligned}
T_8^* &= (\Lambda^7 T_7^*)^{-1/4} \otimes (T_7^* \oplus \Lambda^7 T_7^* ) \  \\
{\bf 8} &=  {\bf 7} \oplus {\bf 1} \\
 \rho_a & =(\rho_m, \rho_8)
\end{aligned}\end{equation}
where $\rho$ is some one-form, and $a=1,...,8$, $m=1,...,7$. Note that the eight-form $\rho_1 \wedge \dots \wedge \rho_8$ is just one on $T_8$. This fits nicely with the fact that only $SL(8)$ acts on this bundle.
This implies that the fundamental ${\bf 56}$ representation (\ref{decomposition_sl8_fund}) decomposes as
\begin{equation}\label{decomposition_gl7_fund}\begin{aligned}
 E_M &= (\Lambda^7 T_7)^{1/2} \otimes (T_7 \oplus \Lambda^2 T_7^* \oplus \Lambda^5 T_7^* \oplus (T_7^* \otimes \Lambda^7 T_7^*)) \ , \\
 {\bf 56} &=  {\bf 7} \oplus {\bf 21} \oplus {\bf 21} \oplus {\bf 7}
\end{aligned}\end{equation}
corresponding respectively to momentum, M2- and M5-brane charge, and KK monopole charge.

In turn, the adjoint \eqref{decomposition_sl8_ad} decomposes into
\begin{equation}\begin{aligned}
A&= T_7 \otimes T_7^* \oplus \Lambda^6 T_7 \oplus \Lambda^6 T_7^* \oplus \Lambda^3T_7 \oplus \Lambda^3 T_7^* \ , \\
{\bf 133} &=  {\bf 48} \oplus {\bf 1} \oplus {\bf 7} \oplus {\bf 7} \oplus {\bf 35} \oplus {\bf 35}
\end{aligned}\end{equation}
We recognize here the $GL(7)$ adjoint (first term), and the shifts of the M-theory 3-form potential $A_3$ (last term) and its 6-form dual $A_6$ (fourth term). These build up the geometric subgroup of transformations that are used to patch the exceptional tangent bundle.
The other pieces correspond to ``hidden symmetries", very much like the $\beta$-transformations in generalized geometry.
Note that $A_3$ and $A_6$ come respectively from the ${\bf 70}$ and ${\bf 63}$ representations of $SL(8)$ in (\ref{decomposition_sl8_ad}), i.e. they embed into the $SL(8)$ representations in \eqref{decomposition_sl8_ad} as
\begin{equation} \label{AshiftsSL8}
A_{\rm{shifts}}=\left( \hat A \otimes \rho_8 , A_4 \right) \ ,
\end{equation}
where we defined the four-form
\begin{equation} \label{A4}
A_4\equiv A_3 \wedge \rho_8 \ ,
\end{equation}
and the vector
\begin{equation} \label{hatA}
\hat A \equiv ({\rm vol}_7)^{-1} \llcorner A_6 \ ,
\end{equation}
with $\llcorner $ meaning the full contraction of $A_6$ with $({\rm vol}_7)^{-1}$.


\section{Supersymmetric reductions of M-theory in EGG}
\label{sec:SUnstr}
In this section we review reductions of M-theory preserving ${N}=2$ and $N=1$ supersymmetry in the language of Exceptional Generalized Geometry following \cite{Grana:2009im,Pacheco:2008ps}. In the $N=2$ case, we show how an $SU(4)$ structure on $T_8$ emerges, and write the EGG structures in terms of the complex and symplectic structures on this space.

To get a supersymmetric effective four-dimensional theory, there should exist  nowhere-vanishing spinors $\eta_i$ on the seven-dimensional internal space such that the eleven dimensional spinor can be decomposed as (for spinor conventions, see \cite{Pacheco:2008ps})
\beq
\epsilon=\xi^i_+ \otimes \eta_i^c + \xi^i_- \otimes \eta_i \ ,
\eeq
where $\xi^i_\pm$ are four-dimensional Weyl spinors, $\eta_i$ are complex $Spin(7)$ spinor and $i=1,.., N$ determines the amount of 4D supersymmetry. When we embed $Spin(7)$ into $Spin(8)$, we can choose $\eta_i$ to be of positive chirality with respect to $Spin(8)$, which means that we can take $\eta_i$ to transform in the ${\bf 8}$ of SU(8). Note that there exists a Majorana condition for $Spin(7)$ (and also for $Spin(8)$) so that actually the real and imaginary parts of $\eta_i$ are independent spinors. Both the real and imaginary part of $\eta_i$ are stabilized by a $Spin(7)$ subgroup in $SL(8,\mathbb{R})$, respectively. Therefore, inside $SL(8,\mathbb{R})$ each complex spinor $\eta_i$ defines a pair of $Spin(7)$ structures. Inside $E_{7(7)}$ though, each $\eta$ transforms in the fundamental of $SU(8)$ and its real and imaginary parts are not independent any more, as they transform into each other under $SU(8)$. Consequently, a single $\eta$ defines an $SU(7)$ structure \cite{Pacheco:2008ps}, and in general $N$ non-mutually parallel spinors define an $SU(8-N) \subset \Es7$ structure.

\subsection{$N=2$ reductions and $SU(6)\times SU(6)$ structures}

For reductions with $N=2$ supersymmetry in four-dimensions, there should be a pair of globally defined no-where vanishing (and nowhere parallel) $SU(8)$ spinors $\eta_1, \eta_2$. As explained above, each of these spinors can be complex so that $\Re \eta_i$ and $\Im \eta_i$ each define an $SU(6)$ structure. Without loss of generality, we can take them to be orthogonal and having the same norm, namely
\begin{equation} \label{normspinSU6}
\bar \eta_i \eta_j=  \e^{-K_L/2} \delta_{ij} \ ,
\end{equation}
where $i,j=1,2$ is an $SU(2)_R$ index, and we introduced an arbitrary normalization factor $\e^{-K_L/2}$ which we will further discuss below.
On the other hand, we have in general the following inner products
\begin{equation} \label{spinSU6generalC}
\eta^T_i \eta_j=  c_{ij} \ ,
\end{equation}
for some complex $2\times 2$ matrix $C=(c_{ij})$.\footnote{The matrix $C$ is complex symmetric and in general not diagonalizable.}

The scalar degrees of freedom of $N=2$ theories, coming from vector and hypermultiplets, are encoded respectively in an $SU(2)_R$ singlet and an $SU(2)_R$ triplet of bispinors \cite{Grana:2009im}. The former embeds in the ${\bf 28}$ representation of $SU(8)$, which appears in the fundamental ${\bf 56}$ representation of $\E7$, and in terms of the decomposition \eqref{decomposition_sl8_fund} reads
\begin{equation}\label{L0}
 \hat L^{(0)} = \tfrac12 (\epsilon^{ij}\eta_i \otimes \eta_j , \epsilon^{ij} \bar \eta_i \otimes \bar \eta_j) \ , \qquad \tilde L^{(0)} = \tfrac12 (-\iu \epsilon^{ij}\eta_i \otimes \eta_j , \iu \epsilon^{ij} \bar \eta_i \otimes \bar \eta_j) \ ,
\end{equation}
where for later convenience we have defined two real bispinors in the ${\bf 28}$ and ${\bf \bar{28}}$ that are related to $L$ by $L = \hat L^{(0)} + \iu \tilde L^{(0)}$. From $\hat L^{(0)} $ we can define also the almost complex structure $J_L$ that relates real and imaginary parts of $L$, given by
\begin{equation}\label{JL}
J_L=2 \, \hat L \times \hat L = L \times \bar L = (\e^{-K_L/2} \delta_{ij} \eta_i \otimes \bar \eta_j - \tfrac14 \e^{-K_L} {\mathbb1 } , 0) \ .
\end{equation}
where ${\mathbb 1}$ is the identity matrix with $Spin(8)$ spinorial indices, $\delta^{\alpha}_{\beta}$.

Furthermore, the $SU(2)_R$ triplet transforms in the ${\bf 63}$ adjoint representation of $SU(8)$, which is embedded in the ${\bf 133}$ adjoint representation of $\E7$, and reads in terms of the decomposition \eqref{decomposition_sl8_ad}
\begin{equation} \label{K0}
 K_a^{(0)} = (\kappa\, \e^{K_L/2} \sigma_{a \ k}^{\ i}\delta^{kj}\eta_i \otimes \bar \eta_j, 0) \
\end{equation}
where $\sigma_a$ are the Pauli sigma matrices and we introduced another normalization factor $\kappa$.
Note that the product of $L$ and $K$ in the ${\bf 56}$ vanishes, i.e.\
\begin{equation}\label{KL_compatibility}
  K_a^{(0)} \cdot L^{(0)} = 0 \ ,
\end{equation}
which means that the stabilizers of $L^{(0)}$ ($\Ex6\subset \Es7$) and $K_a^{(0)}$ ($SO^*(12) \subset \Es7$) intersect in $SU(6)$.
General $L$ and $K_a$ are then constructed by acting with the shift matrix $A_{\rm shifts}$ in the adjoint of $\Es7$ on them, i.e.\
\begin{equation}\label{LK_Ashifts}
L= \e^{A_{\rm shifts}} L^{(0)}   \ , \qquad    J_L= \e^{A_{\rm shifts}} J_L^{(0)}   \ , \qquad    K_a = \e^{A_{\rm shifts}} K_a^{(0)} \ .
\end{equation}

The K\"ahler potential for the space of structures $L$ and the hyper-K\"ahler potential hyper-K\"ahler cone over the space of $K_a$ has been given in \cite{Grana:2009im}. The K\"ahler potential for $L$  is given by the moment map for the rotation of $L$ by a phase (generated by $J_L$) and can be expressed by the logarithm of the quartic invariant of $\hat L$ \cite{Grana:2009im}
\begin{equation}\label{KahlerL}
K_L = -\log \sqrt{-\tfrac14 (J_L,J_L)}=- \log \sqrt{- q(\hat L)} =- \log \left(- \tfrac{\iu}{8} \langle L^{(0)}, \bar L^{(0)} \rangle \right) \ .
\end{equation}
where $( \cdot, \cdot)$ refers to the trace in the adjoint, given in (\ref{tradj}). The hyper-K\"ahler potential of the hyper-K\"ahler cone over the moduli space of $K_a$ is determined by \cite{Grana:2009im}
\begin{equation}\label{KahlerK}
\kappa =  \sqrt{-\tfrac14 (K_a,K_a)} \ .
\end{equation}
It determines the normalization of the $K_a$ by
\begin{equation}
K_a\cdot (K_b \cdot \nu) = - \tfrac12 \kappa^2 \delta_{ab} \nu + \kappa \, \epsilon_{abc} K_c \cdot \nu \ ,
\end{equation}
where $\nu$ is an arbitrary element in the $56$ representation.

The supersymmetric couplings between the two objects $L$ and $K_a$ are given by the Killing prepotentials \cite{Grana:2009im,Waldram:2012}
\begin{equation}
 P^a = \epsilon^{abc} ( D_L K_b, K_c ) \ ,
\end{equation}
and  $D_L = \langle L , D\rangle - L \times D$ is the Dorfman derivative \cite{Coimbra:2011ky} along $L$, with $D$ being the standard differential operator.\footnote{$\langle L,D \rangle$ is the symplectic invariant (\ref{symplinvt}), and $(L \times D)$ is the projection onto the adjoint in the product of two {\bf 56} elements, given in (\ref{56x56=133}).}
Inserting this, we get \cite{Waldram:2012}
\begin{equation}\label{prepotentials}
 P^a = \epsilon^{abc} ( \langle L, D \rangle K_b, K_c ) + 4 \kappa( L, D K_a ) \ .
\end{equation}

\subsection{$N=2$ reductions and $SU(4)$ structure on $T_8$}
In the last section we defined a general $SU(6)$ structure with an arbitrary product \eqref{spinSU6generalC}. Note that in general, the spinors $\eta_i$ are complex, and real and imaginary part of each spinor, if they are never parallel, already define an $SU(6)$ structure, so in general we would have ``$SU(6) \times SU(6)$'' structures (and secretly, a theory with more supersymmetries). In such a generic case, it is difficult to be more explicit as typically all $SL(8)$ components are present  in $L$ and $K_a$. The case of real (i.e.\ Majorana) spinors $\eta_i$ simplifies the form of $L$ and $K_a$ and gives rise to a natural interpretation in terms of the bundle $T_8$.
The relation \eqref{spinSU6generalC} reduces for real spinors to the normalization condition \eqref{normspinSU6} so that
\begin{equation}
c_{ij} = \e^{-K_L/2}\delta_{ij} \ .
\end{equation}

We will express now both $L^{(0)}$ and $K_a^{(0)}$ in \eqref{L0} and \eqref{K0} in terms of objects in $SL(8,\mathbb{R})$ representations. For this we define a pure spinor
\beq
\chi = \eta_1 + \iu \eta_2
\eeq
out of the real spinors $\eta_i$.\footnote{A complex $Spin(8)$ spinor $\chi$ is called pure if $\chi^T \chi = 0$.}
We then find for $L^{(0)}$ and $K_a^{(0)}$ that
\begin{equation} \label{adjointchi}\begin{aligned}
  K_3^{(0)} \pm \iu K_1^{(0)} = &(\chi \otimes \chi^T , 0) \ , \\
  J_L^{(0)} \pm \iu K_2^{(0)} = &(\chi \otimes \bar \chi - \tfrac14 \e^{K_L/2} {\mathbb 1} , 0) \ .
\end{aligned} \end{equation}
On the other hand, the bispinors $\Phi_1 = \chi \otimes \bar \chi$ and $\Phi_2 = \chi \otimes \chi^T$ are a pair of compatible $O(8,8)$ pure spinors that define an $SU(4)$ structure on $T_8$, given by \cite{Hitchin}
\begin{equation} \label{puresp}
\Phi_1 = \tfrac14 \e^{-K_L/2} {\rm exp}(-\iu \e^{K_L/2}J^{(0)}_8) \ , \qquad \Phi_2 = \Omega^{(0)}_4 \ ,
\end{equation}
where we made the following definitions
\begin{equation}\begin{aligned}
 J_8^{(0)} = &-\iu \,   \bar \chi \gamma_{ab} \chi \, \diff x^a \wedge \diff x^b =\epsilon^{ij}  \bar \eta_i \gamma_{ab} \eta_j \diff x^a \wedge \diff x^b \ , \\
\Omega_4^{(0)}=& -\tfrac{1}{4} \kappa\, \e^{K_L/2}  \chi^T \gamma_{abcd} \chi  \, \diff x^a \wedge \diff x^b \wedge \diff x^c \wedge \diff x^d \\
=& -\tfrac{1}{4} \kappa\, \e^{K_L/2}  (\bar \eta_1 + \iu \bar \eta_2) \gamma_{abcd} (\eta_1 + \iu \eta_2)  \diff x^a \wedge \diff x^b \wedge \diff x^c \wedge \diff x^d \ .
\end{aligned}\end{equation}
In terms of the $x^m$ we find
\begin{eqnarray}
&J_8^{(0)} = J + v^{(0)} \wedge \rho_8 \ , \qquad J =  \epsilon^{ij}  \bar \eta_i \gamma_{mn} \eta_j \diff x^m \wedge \diff x^n \ , \\
 &\Omega_4^{(0)} = (\rho_8+\iu v^{(0)}) \wedge \Omega \ , \qquad
\Omega = \iota_{\hat \rho_8} \Omega_4^{(0)}  \ ,
\end{eqnarray}
where the one-form $v^{(0)}$ is defined by\footnote{The hat denotes the vector dual to a one-form. In the special case that $v$ is the one-form $\diff y$ of a circle, one can reduce the theory to type IIA, where $v_m \gamma^{m8}$ then plays the role of chirality.}
\begin{equation}
 v^{(0)} = \iota_{\hat \rho_8} \epsilon^{ij}  \bar \eta_i \gamma_{ab} \eta_j \diff x^a \wedge dx^b  \ .
\end{equation}
The four-form $\Omega_4^{(0)}$ defines a complex structure $I_8^{(0)}$ by
\begin{equation}\begin{aligned} \label{I80}
(I_8^{(0)})^m{}_n=& \tfrac{1}{\kappa^2} (*_8 \Im\Omega^{(0)}_4)^{mpqr} (\Re\Omega^{(0)}_4)_{npqr}
\\ = & \tfrac{1}{2\kappa^2} ((\hat v^{(0)})^m (\rho_8)_n - \hat \rho_8^m v^{(0)}_n + (\Im \Omega)^{mpq} (\Re \Omega)_{npq} - (\Re \Omega)^{mpq} (\Im \Omega)_{npq}) \   .
\end{aligned}\end{equation}
In terms of these objects, the structure $L^{(0)}$ has the following $\SLE$ decomposition (cf.~\eqref{decomposition_sl8_fund})\footnote{For the generic case of complex spinors $\eta_i$, the form of $L^{(0)}$ is also given by Eq.\ (\ref{L0SU6}), but in this case $J_8^{(0)}$ is complex, given by $$J_8^{(0)}=\epsilon^{ij}  \eta^T_i \gamma_{ab} \eta_j \diff x^a \wedge \diff x^b \ ,$$ and similarly for $J$ and $v^{(0)}$.}    
\begin{equation}\label{L0SU6}\begin{aligned}
 L^{(0)}= \hat L^{(0)}+ \iu \tilde L^{(0)} = & (- \hat \rho_8 \wedge \hat v^{(0)} + \e^{K_L} \vol^{-1}_8 \llcorner (\tfrac12 \rho_8 \wedge v^{(0)}\wedge  J\wedge J) ,\ \iu  (\rho_8 \wedge v^{(0)} -J))\\ = & (\e^{K_L} \vol^{-1}_8 \llcorner (\tfrac{1}{3!}J_8^{(0)} \wedge J_8^{(0)} \wedge J_8^{(0)}),\ -\iu J_8^{(0)}) \\ = & \e^{\iu \rho_8\wedge v^{(0)}\wedge J} \cdot (-\hat \rho_8 \wedge \hat v^{(0)},\ 0)\ ,
\end{aligned}\end{equation}
where we have used \eqref{fundsu8tosl8} and $\vol^{-1}_8 \llcorner $ means contraction with the eight-dimensional volume form, i.e.\ with the numeric epsilon tensor (the $8$-dimensional space has no volume modulus).
We see that $ L^{(0)}$ can be understood as the shift of the bi-vector $-\hat \rho_8 \wedge \hat v$ by some imaginary four-form $\iu \rho_8 \wedge v^{(0)} \wedge J$.\footnote{Note that this shift is not an $E_{7(7)}$ transformation because the four-form is not real.}
The almost complex structure relating real and imaginary part of $L^{(0)}$ given in \eqref{JL} can be computed to be
\begin{equation}
 J_{L^{(0)}}= (0 , J_8 \wedge J_8) \ .
\end{equation}
Similarly, using  \eqref{adjointchi} and \eqref{pw2}, we can determine $K_a^{(0)}$ for the decomposition \eqref{decomposition_sl8_ad} to be
\begin{equation}\label{K0SU6}\begin{aligned}
 K_1^{(0)} = & (0, \Im \Omega^{(0)}_4 ) \ ,\\
 K_2^{(0)} = & (\kappa I^{(0)}_8, 0)\ ,\\
 K_3^{(0)} = & (0, \Re \Omega^{(0)}_4 ) \ .
\end{aligned}\end{equation}

From the Fierz identities (or from the $SU(2)$ algebra) it follows that $\Omega^{(0)}_4$ transforms holomorphically under $I_8^{(0)}$, and obeys
\begin{equation}
\Omega_4^{(0)} \wedge \Omega^{(0)}_4 = 0 \ , \qquad \Omega^{(0)}_4 \wedge \bar \Omega^{(0)}_4= 2 \kappa^2 \,  {\rm vol_8} \ .
\end{equation}
 We also have \ $\iota_{\hat v^{(0)}} J=\iota_{\hat \rho_8} J=0$, and  $\tfrac{1}{3!} v \wedge J \wedge J \wedge J = \e^{-2K_L}\vol_7$ or in other words
\begin{equation}
\tfrac{1}{4!} J^{(0)}_8 \wedge J^{(0)}_8 \wedge J^{(0)}_8 \wedge J^{(0)}_8 = \tfrac{1}{3!}  J \wedge  J \wedge J \wedge v \wedge \rho_8= \e^{-2 K_L} \rm{vol_8} \ .
\end{equation}
Furthermore, the compatibility condition \eqref{KL_compatibility} of $L^{(0)}$ and $K_a^{(0)}$ implies $\iota_{\hat v^{(0)}} \Omega = 0$ and $J \wedge \Omega = 0$, or in other words
\begin{equation}
 J^{(0)}_8  \wedge \Omega^{(0)}_4=0 \ .
\end{equation}
Therefore $J^{(0)}_8$ and $\Omega^{(0)}_4$ define an $SU(4)$ structure on the eight-dimensional space $T_8$, i.e.\ inside $SL(8,\mathbb{R})$.\footnote{As stated above, these objects define an $SU(6)$ structure inside $E_{7(7)}$.}
From the compatibility condition \eqref{KL_compatibility} it also follows that $g^{(0)}_8$ defined by
\begin{equation} \label{metricJI}
  g^{(0)}_8 = \e^{K_L/2} J^{(0)}_8 \cdot I_8^{(0)} \ ,
\end{equation}
is a symmetric object. Furthermore, explicit computation shows that $g^{(0)}_8$ is of the form
\begin{equation} \label{metric8}
g_8^{(0)}=({\rm det  g_7})^{-1/4} \left(\begin{array}{cc} g_7 & 0 \\ 0 & {\rm det} g_7 \end{array} \right) \ .
\end{equation}
Similar to \eqref{metricJI} we also find for the inverse metric $g^{(0)\, -1}_8$ the expression
\begin{equation} \label{metricinverseJI}
 g^{(0)\, -1}_8 = \e^{3K_L/2} I_8^{(0)} \cdot  (\vol^{-1}_8 \llcorner  (\tfrac{1}{3!} J^{(0)}_8 \wedge J^{(0)}_8 \wedge J^{(0)}_8)) \ .
\end{equation}

General elements $L$, $K_a$ in the orbit can be achieved from $L^{(0)}$, $K_a^{(0)}$ by the action of $E_{7(7)}$, cf.\ \eqref{LK_Ashifts}. The degrees of freedom that can modify the above embeddings are the remaining massless fields of M-theory, namely the three-form gauge field $A_3$ and its magnetic dual six-form field $A_6$. Their action in $SL(8)$ language is shown in \eqref{AshiftsSL8}.
This gives
\begin{equation}\label{LSU6}\begin{aligned}
L  = & \e^{A_4} (\vol^{-1}_8 \llcorner ( \tfrac{1}{3!} J_8 \wedge J_8 \wedge J_8),\ -\iu J_8)  \ , \\
J_L  = & -\e^{A_4} (0,\ J_8 \wedge J_8)  \ ,
\end{aligned}\end{equation}
and
\begin{equation}\label{KSU6}\begin{aligned}
 K_1 = & \e^{A_4} (0, \Im \Omega_4)\ ,\\
 K_2 = & \e^{A_4} (\kappa I_8, 0)\ ,\\
 K_3 = & \e^{A_4} (0, \Re \Omega_4) \ ,
\end{aligned}\end{equation}
where $A_4$ is defined in (\ref{A4}) and the shifted $J$ and $\Omega_4$ are
\begin{equation} \label{Omega4A6}
J_8 = J + v \wedge \rho_8 \ ,  \qquad \Omega_4=\left( (1+\iu a_6^v) \rho_8+\iu v^{(0)} \right) \wedge \Omega - \iu \rho_8 \wedge v^{(0)} \wedge \iota_{\tilde A} \Omega
\end{equation}
with $\hat A$ defined in (\ref{hatA}), and
\begin{equation}
  v = v^{(0)} - \iota_{\hat A} J \ ,\qquad a_6^v= \vol_7^{-1} \cdot (v^{(0)} \wedge A_6) \ .
\end{equation}
The symplectic and complex structure  $J_8$ and $\Omega_4$ are also compatible, which means $J_8  \wedge \Omega_4=0$.
Finally, the almost complex structure $I^{(0)}_8$ is shifted so that $\Omega^{(0)}_4$ is replaced by $\Omega_4$ in \eqref{I80}.

We see that the description of an $SU(6) \subset E_7$ structure in M-theory is completely analogous to the type II case discussed in \cite{Grana:2009im}, namely given by one object in the fundamental representation and a triplet in the adjoint. These are in turn described respectively by $J_8$ and $\Omega_4$, which define an $SU(4) \subset SL(8)$ structure, in a form that very much resembles the pure spinors $e^{-\iu J_8}$, $\Omega_4$ of two generalized almost complex structures. We will come back to this in Section~\ref{sec:IIA}.


We turn now to the K\"ahler potentials and prepotentials for the space of structures $L$ and (the hyper-K\"ahler cone over) $K_a$. They both have the expected form in terms of the $SU(4)$ structure objects $J_8$ and $\Omega_4$.
The K\"ahler potential for $L$ can be easily computed from \eqref{KahlerL}to be
\begin{equation}
K_L = - \tfrac12 \log(\tfrac{1}{4!} J_8 \wedge J_8 \wedge J_8 \wedge J_8)  \ .
\end{equation}
The hyper-K\"ahler potential of the hyper-K\"ahler cone over the moduli space of $K_a$ is given by \eqref{KahlerK} and reads in terms of $\Omega_4$ as
\begin{equation}
\kappa = \sqrt{\tfrac12 \Omega_4 \wedge \bar \Omega_4} \ .
\end{equation}
Finally, the Killing prepotentials \eqref{prepotentials} are computed in Appendix \ref{appprepotential}, using the differential operator $D$ in the ${\bf 56}$ representation that is given by the embedding
\begin{equation}\label{diff_op}
D = ( 0 , \rho_8 \wedge \diff) \ .
\end{equation}
The result of the computation is given by
\begin{equation}\label{prepotentials_sol}\begin{aligned}
 P^3+ \iu P^1 =\ & -4 \kappa \left(\iu  J \wedge \rho_8 \wedge \diff  \Omega^{(0)}_4 +   \iota_v \Omega^{(0)}_4 \wedge \rho_8 \wedge \diff A_3 \right) \ , \\
 P^2 =\ &\e^{-K_L} \Im \left(({\cal L}_{\hat v}   \bar \Omega^{(0)}_4) \wedge \Omega^{(0)}_4\right) - 2 \kappa^2 \e^{K_L} \rho_8 \wedge J \wedge J \wedge  v^{(0)} \wedge \diff v^{(0)} \\ &
 + 4 \kappa \rho_8 \wedge \left( \diff A_6 - (A_3 + \iu \kappa v^{(0)} \wedge J) \wedge   \diff A_3 \right)   \ , 
\end{aligned} \end{equation}
where ${\cal L}_{\hat v}$ is the Lie derivative along $v$.

\subsection{$N=1$ reductions}
\label{sec:SU(7)}

For reductions with $N=1$ supersymmetry, there is a single $Spin(8)$ internal spinor $\eta$, which defines an $SU(7)$ structure in $\Es7$.\footnote{This spinor can have a real and imaginary part and thereby actually defines an $SU(7) \times SU(7)$ structure.} This structure can be encoded in a nowhere-vanishing object $\phi$ in the ${\bf 912}$ representation of $E_{7(7)}$ \cite{Pacheco:2008ps}.
The ${\bf 912}$ decomposes in $SU(8)$ representations in the same way as for $SL(8)$, given in \eqref{rep912}. In terms of this decomposition,
we have
\begin{equation}\label{embedding_spinor_912}
 \phi^{(0)} =  (2 \eta \otimes \eta, 0, 0,0) \ , \quad \bar \phi^{(0)} = (0, 0, 2 \bar \eta \otimes \bar \eta,0) \ ,
\end{equation}
with the normalization
\begin{equation}\label{normspinSU7}
\bar \eta \eta = \e^{-K_\phi/2} \ .
\end{equation}
Using  (\ref{36420}) we can write this in terms of $SL(8)$ representations (\ref{rep912})
\begin{equation}
\phi^{(0)}= - ( c ( g^{(0)}_8)^{-1}, g_8^{(0)} \cdot ({\rm vol}^{-1}_8 \llcorner \phi^{(0)}_4) , \iu c g_8^{(0)} ,\iu (g_8^{(0)})^{-1} \cdot \phi^{(0)}_4)
\end{equation}
where $g_8^{(0)}$ is the metric on $T_8$ (\ref{metric8}), the complex constant $c$ is defined by
\begin{equation}
 \eta^T \eta = c \ ,
\end{equation}
and
\begin{equation}
\phi^{(0)}_{abcd} = \tfrac{3}{8}\eta^{T} \gamma_{abcd} \eta \ .
\end{equation}
Note that at points where $|c| \ne \e^{K_\phi/2}$ real and imaginary part of $\eta$ actually define a (local) $SU(6)$-structure. 
In terms of the split into $7+1$ in (\ref{decomposition_gl7_V}), this can be written as
\begin{equation}
\phi^{(0)}_4=\rho_8 \wedge \alpha_3  + \ast_7 \alpha_3 \ ,
\end{equation}
where $\alpha_3$ is a complex three-form that defines a $G_2\times G_2$-structure which reduces to a real three-form in the $G_2$-structure case (i.e.\ when $\eta$ is Majorana). Note that $\phi^{(0)}_4$ is self-dual as $\eta$ is of positive chirality in $Spin(8)$.

A general element $\phi$ is obtained by
\begin{equation}\label{phi_SU7}
\phi = \e^{A_{\rm shifts}} \phi^{(0)} = - \e^{ A_4}  (  c g_8^{-1}, g_8 \cdot ({{\rm vol}_8^{-1}} \llcorner \phi_4) , \iu c g_8 \, ,\iu g_8^{-1} \cdot \phi_4) \ ,
\end{equation}
where $A_4$, defined in \eqref{A4}, acts by the $E_7$ adjoint action \eqref{912adjoint_sl8}, while $A_6$ shifts the metric (\ref{metric8}) by
\begin{equation} \label{metric8A6}
g_8=({\rm det  g_7})^{-1/4} \left(\begin{array}{cc} g_7 + (\ast_7 A_6)^2  & ({\rm det} g_7)^{1/2} \ast_7 A_6 \\ ({\rm det} g_7)^{1/2} \ast_7 A_6 & {\rm det} g_7 \end{array} \right) \ \ ,
\end{equation}
and the 4-form
\begin{equation}\label{phi4A6}
\phi_4=\phi^{(0)}_4 + \rho_8 \wedge \iota_{\hat A} \ast_7 \alpha_3 \ .
\end{equation}
Note that the volume form ${\rm vol}_8$ of the metric $g_8$ in \eqref{metric8A6} is still given by the eight-dimensional epsilon tensor (with entries $\pm 1$ and $0$), as there is no $\mathbb{R}_+$ factor in $E_{7(7)}$ corresponding to the eight-dimensional volume. We will therefore sometimes abuse notation and make no distinction between an eight-form and a scalar.


The stabilizer of $\phi$ turns out to be indeed $SU(7)$ \cite{Pacheco:2008ps}. Therefore, the existence of $\phi$ is completely equivalent to $\eta$. We will denote the real and imaginary parts of $\phi$ by $\hat \phi$ and $\tilde \phi$.
The product ${\bf 912} \times {\bf 912} \to {\bf 133}$ gives us the generator  ${\cal J}_{\phi}$ for the almost complex structure $\e^{{\cal J}_{\phi}}$ related to $\phi$. More precisely, we have
\begin{equation}
{\cal J}_{\phi} = 2 \hat \phi \times \hat \phi \ ,
\end{equation}
which in terms of the $SL(8)$ decomposition (\ref{decomposition_sl8_ad}) is
\begin{equation}\label{cs_SU7}
{\cal J}_{\phi} = \e^{A_4}\cdot ( I_\phi , \phi^r_4) ) \ ,
\end{equation}
with
\begin{equation}
(I_\phi)^a{}_b = -\tfrac43  (\vol^{-1}_8 \llcorner \Re \phi_4)^{acde} (\Im \phi_4)_{bcde}
\end{equation}
and
\begin{equation}
 \phi^r_4 = 14 \Re(\bar c \phi_4) = \tfrac{1}{2} \e^{K_\phi/2} \bar \eta \gamma_{abcd} \eta \, \diff x^a \wedge \diff x^b \wedge \diff x^c \wedge \diff x^d \ .
\end{equation}
The K\"ahler potential $K_\phi$ in the orbit of $\phi$ is given by the quartic invariant in the ${\bf 912}$ and turns out to be
\begin{equation}\label{Kahlerpot_SU7}\begin{aligned}
K_\phi = & - \log \sqrt{-q(\hat \phi)} = - \log \sqrt{-\tfrac14 ({\cal J}_\phi,{\cal J}_\phi)}   = -\log \sqrt{-\tfrac14 \tr (I_\phi^2)} -\tfrac12 \log \tfrac{1}{4!} \phi^r_4 \wedge \phi^r_4  \\   = & -\log \sqrt{-\tfrac14 \tr ((I^{(0)}_\phi)^2)} -\tfrac12 \log \tfrac{1}{4!} \phi_4^{r\,(0)} \wedge \phi_4^{r\,(0)}  \ ,
\end{aligned}\end{equation}
where in the last step we used $(\iota_{\hat A} \ast_7 \alpha_3) \wedge \ast_7 \alpha_3=0$. Therefore, the K\"ahler potential does not depend on $A_6$ and $A_4$, in agreement with $E_{7(7)}$ invariance.
Note that this expression reduces to the known one in the $G_2$-structure case.

As has been derived in \cite{Pacheco:2008ps}, the superpotential $W$ is given through the eigenvalue equation
\begin{equation} \label{def_superpotential}
(D \phi) \cdot \phi = (\e^{K_\phi/2} W) \phi \ ,
\end{equation}
where $D$ is defined in (\ref{diff_op}). We present the detailed computation of the superpotential in Appendix \ref{appsuperpotential}.  We get
\begin{equation}\label{superpotential_SU7}
 \e^{K_\phi/2}W = \iota_{\rho_8} \phi_4^{(0)} \wedge \diff \phi_4^{(0)} + \rho_8 \wedge \diff A_6 -2 \iu \phi_4 \wedge \iota_{\rho_8} \diff A_4 + \iota_{\rho_8} A_4 \wedge \diff A_4 \ .
\end{equation}
which is exactly the result of \cite{Pacheco:2008ps}. A similar result for manifolds of $G_2$ structure has been obtained in \cite{House:2004pm}.

\subsection{$N=2$ in the language of $N=1$}

To finish this section, we note that the $SU(6)$ structure of $N=2$ reductions can also be understood as a triple of $SU(7)$ structures, by using the product ${\bf 133}\times {\bf 56} \to {\bf 912}$ given in \eqref{cross_product_sl8}. We find the objects
\begin{equation}\label{phi_KL}\begin{aligned}
 \hat \phi^{(0)}_a =  \hat L^{(0)} \otimes K_a^{(0)} = 2\kappa(\epsilon_{ik}(\sigma_{a})^{k}{}_{j}\eta_i \otimes \eta_j, 0, \epsilon_{ik}(\sigma_{a})^{k}{}_{j} \bar \eta_i \otimes \bar \eta_j ,0) \ ,\\
 \tilde \phi^{(0)}_a =  \tilde L^{(0)} \otimes K_a^{(0)} = 2 \kappa (-\iu \epsilon_{ik}(\sigma_{a})^{k}{}_{j}\eta_i \otimes \eta_j, 0, \iu(\epsilon_{ik}(\sigma_{a})^{k}{}_{j}\bar \eta_i \otimes \bar \eta_j ,0) \ ,
\end{aligned}\end{equation}
which are non-zero and define an $SU(2)$ triplet of $SU(7)$ structures of the form given in \eqref{embedding_spinor_912}.
The index $a$ here labels the various symmetric combinations of the $\eta_i$. A spinor $\eta = \alpha_i \eta_i$ corresponds then to the $SU(7)$ structure defined by $\phi= (\alpha_i \epsilon_{ik}(\sigma_{a})^{k}{}_{j} \alpha_j) \phi_a$, where $\phi_a^{(0)}=\hat \phi_a^{(0)} + \iu \tilde \phi_a^{(0)}$.

For an $SU(6)$ structure the $\phi_a^{(0)}$ can be computed in terms of the $SU(4)$ forms $J_8$ and $\Omega_4$ on the 8-dimensional bundle. For this, we use \eqref{metricJI} and \eqref{metricinverseJI}. In terms of the decomposition \eqref{rep912}, we then find
\begin{equation} \begin{aligned}
\phi_1^{(0)}  = & - 2 \e^{-K_L/2} (0,\ \iu g^{(0)}_8 \cdot (\vol^{-1}_8 \llcorner  \Re \Omega^{(0)}_4),\ 0 ,\ (g^{(0)}_8)^{-1} \cdot \Re \Omega^{(0)}_4 ) \ , \\
\phi_2^{(0)}  = & - 2 \e^{-K_L/2} \kappa(( g^{(0)}_8)^{-1},\ \e^{K_L} g^{(0)}_8 \cdot (\vol^{-1}_8 \llcorner  (J_8^{(0)} \wedge J_8^{(0)})) ,\\ & \qquad\qquad\qquad \iu g^{(0)}_8 ,\ \iu \e^{K_L} (g^{(0)}_8)^{-1} \cdot (J_8^{(0)} \wedge J_8^{(0)}) )   \ ,\\
\phi_3^{(0)}   = & -2\e^{-K_L/2}  (0,\ -\iu g^{(0)}_8 \cdot (\vol^{-1}_8 \llcorner \Im \Omega^{(0)}_4),\ 0 ,\ (g^{(0)}_8)^{-1} \cdot \Im \Omega^{(0)}_4 )   \ ,
\end{aligned} \end{equation}
where $g^{(0)}_8$ is given in (\ref{metric8}). As before, the general form of $\phi_a $ is obtained by acting with $A_6$ and $A_3$ on the above expressions. The action of $A_6$ turns $g_8^{(0)}$ into $g_8$ (cf.\ Eq.\ \eqref{metric8A6}), $J_8^{(0)}$ into $J_8$ and $\Omega_4^{(0)}$ into $\Omega_4$ (see \eqref{Omega4A6}), while $A_3$ acts as usual by the 4-form shift $\e^{\rho_8 \wedge A_3}$.

One can therefore understand the $SU(6)$ structure in $E_7$ relevant for $N=2$ compactifications as a triple of $SU(7)$ structures. When we perform a projection on the $SU(6)$ structure, for instance by orbifolding or orientifolding, a single combination of these $SU(7)$ structures will survive, giving the $N=1$ description one expects. We will discuss such projections in Section \ref{sec:Orbifolding}.

\section{Type IIA reductions}
\label{sec:IIA}
\subsection{Type IIA and $GL(6)$ decompositions}

To descend to type IIA, we further split the 7-dimensional tangent space into six plus one-dimensional pieces, i.e.
\bea
T^*_7 &=&T^*_6 \oplus T^*_y \ \\
\rho_m&=&(\rho_{\hm}, \rho_y) \nonumber
\eea
Choosing $T_y$ to have no $GL(6)$ weight, we get that the exceptional tangent bundle
(\ref{decomposition_gl7_fund}) decomposes as ${\bf 56} =  {\bf 6} \oplus {\bf 6} \oplus {\bf 6} \oplus {\bf 6} \oplus {\bf 32}$, i.e.
\begin{equation} \begin{aligned}
 E_{\rm IIA} = & (\Lambda^6 T_6)^{1/2} \otimes \left((T_6 \oplus \Lambda_0 T^*_6 \oplus \Lambda^2 T^*_6 \oplus T^*_6 \oplus \Lambda^5 T^*_6  \oplus \Lambda^4 T_6^* \oplus (T_6^* \otimes \Lambda^6 T^*) \oplus \Lambda^6 T^* \right)  \\
 =& (\Lambda^6 T_6)^{1/2} \otimes \left((T_6 \oplus T^*_6 \oplus \Lambda^5 T^*_6  \oplus (T_6^* \otimes \Lambda^6 T^*) \oplus \Lambda^{\rm{even}}T^*_6 \right) \ ,
\end{aligned}\end{equation}
where from the last expression we recognize the type IIA charges, namely momentum, winding, their magnetic duals and the p-brane charges.

The adjoint decomposes as
\begin{equation}\begin{aligned}
  A_{\rm IIA} = & [\lambda \oplus (T_6 \otimes T^*_6)_0 \oplus (T_y \otimes \Lambda^2 T_6)\oplus (T^*_y \otimes \Lambda^2 T^*_6)] \oplus [\varphi \oplus \Lambda^6 T_6 \oplus \Lambda^6 T^*_6] \\ & \oplus [ (T_y \otimes T^*_6 \oplus \Lambda^3 T^*_6 \oplus (T^*_y \otimes \Lambda^5 T^*_6)] \oplus [ (T^*_y \otimes T_6 \oplus \Lambda^3 T_6 \oplus (T_y \otimes \Lambda^5 T_6)] \ \\
  =& \left(T_6\otimes T^*_6\right)
            \oplus \Lambda^2T_6 \oplus \Lambda^2T^*_6
            \oplus \bbR \oplus \Lambda^6T^*_6 \oplus \Lambda^6T_6
            \oplus \Lambda^\textrm{odd}T^*_6
            \oplus \Lambda^\textrm{odd}T_6
\end{aligned}\end{equation}
where
\begin{equation}
\lambda = \diff y \otimes \partial_y \ , \quad \varphi = 2 \diff y \otimes \partial_y + \diff x^{\hat m} \otimes \partial_{\hat m} \ ,
\end{equation}
are linear combinations of the generators for shifts in the volumes of the six-dimensional part and the eleventh dimension. From the last equality we recognize the $O(6,6)$ adjoint, where the $GL(6)$ and $B_2$-field transformations live, the $SL(2)$ adjoint where the dual $B_6$ of $B_2$ is, and the sum of odd forms corresponding to the shifts of the RR potentials.

More precisely, under the breaking of the U-duality group $E_{7(7)}$ into the product of T-duality and S-duality group $SO(6,6) \times SL(2)$ ,the fundamental representation splits according to\footnote{\label{foot:S-duality}$S$-duality here does not refer to the type IIB S-duality, but as the group that acts by fractional linear transformations on $\tau=B_6 + \iu e^{-\phi}$.}
\begin{equation} \label{56Tsub} \begin{array}{lc}
{\bf 56}  =& ({\bf 12},{\bf 2}) \oplus ({\bf 32},{\bf 1})  \ , \\
L^{\cal A} =& (L^{Ai} \ , \  L^\pm)
\end{array}
\end{equation}
where $A=1,...,12$ ($i=1,2$) is a fundamental $O(6,6)$ ($SL(2)$) index, and $\pm$ represents a positive or negative chirality $O(6,6)$ spinor (the plus is relevant for type IIA and M-theory, while for type IIB we need the negative chirality representation). The adjoint decomposes as the adjoint of each group, plus a spinor contribution
\begin{equation} \label{133Tsub} \begin{array}{lc}
{\bf 133} =& ({\bf 66},{\bf 1}) \oplus ({\bf 1},{\bf 3}) \oplus ({\bf 32'},{\bf 2}) \ , \\
K=& (\ K^A{}_B \ \ ,  \ K^i{}_j \ , \  \ K^{\mp i}) \ .
\end{array}
\end{equation}
The embedding of the gauge fields $B_2, B_6$ and $C^\mp$ (that we will call collectively $A_{\rm{shifts}}$) in type IIA and type IIB is the following\cite{Grana:2009im}
\beq \label{AshiftsTsub}
A_{\rm{shifts}}=\left( \left(\begin{array}{cc} 0 & B_{2} \\ 0 & 0 \end{array}\right), \left(\begin{array}{cc} 0 & B_{6} \\ 0 & 0 \end{array}\right),\left(\begin{array}{c} C^{\mp} \\ 0 \end{array}\right)\right) \ .
\eeq

We can now write explicitly the one-form $v$ and gauge fields $A_3$ (or $A_4$), $A_6$ of the previous sections in terms of their type IIA counterparts
\begin{equation}
v^{(0)} = \diff y +C_1 \ , \qquad C_5 = \iota_{\partial_y} A_6 \ , \qquad
A_4 = \rho_8 \wedge (C_3 - \diff y \wedge B_2) \ ,\qquad \iota_{\partial_y} C_3 = 0 \ .
\end{equation}
Inspecting (\ref{AshiftsSL8}), we can write the $\Es7$ embedding of the gauge fields in $\SLE$ decomposition (cf. (\ref{decomposition_sl8_ad})) as follows
\beq \label{AshiftsSL8IIA}
A_{\rm{shifts}}=(C_1\otimes \partial_y + \rho_8 \otimes \hat C_5 \, , \, \rho_8 \wedge C_3 - \rho_8\wedge \diff y \wedge B)
\eeq
where $\hat C_5$ is the vector associated to $C_5$, i.e.
\beq
\hat C_5={\rm vol}_6^{-1} \cdot C_5 \ .
\eeq

\subsection{$N=2$ and $N=1$ reductions}

As in M-theory, reductions to 4D with $N=n$ supersymmetry require $n$ internal $SU(8)$ spinors $\eta^i$, which show up
in the decomposition of the two ten-dimensional spinors $\epsilon^{1,2}$ as follows
\begin{align}
\left(\begin{array}{c}\epsilon^1\\ \epsilon^2\end{array}\right)=\xi^i_-\otimes \eta_i  +\mbox{c.c.}\label{suvar}
\end{align}
where $\eta_i$ are $SU(8)$ spinors that combine the $SU(4)\cong O(6)$ spinors that build up $\epsilon_1$ and $\epsilon_2$.

Given $\eta_{1,2}$ and $\eta$ for the case of $N=2$ and $N=1$, the reductions work exactly as in M-theory, namely one builds the structures $L^{(0)}, K_a^{(0)}$ as in (\ref{L0}), (\ref{K0}) in the case of $N=2$ \cite{Grana:2009im}, and $\phi^{(0)}$ as in (\ref{embedding_spinor_912}) in the case of $N=1$.  Their full orbit is obtained by the $\Es7$ action of the gauge fields (\ref{AshiftsSL8IIA}).

Let us concentrate on the $N=2$ case now. The form of $L$ resembles very much  the pure spinor counterpart $e^{B+\iu J}$ for an $SU(3)$ structure. Indeed, we get
\begin{equation}\label{LSU6_IIA}
L  = \e^{(C_1\otimes \partial_y + \rho_8 \otimes \hat C_5,\rho_8 \wedge C_3)} \e^{\rho_8\wedge \diff y \wedge (-B- \iu J)} \cdot (\hat \rho_8 \wedge \hat v,\ 0) \ .
\end{equation}
Note however that $J$ here is a 2-form constructed as a bilinear of two different $SU(8)$ spinors, and furthermore it is complex.
To make the comparison with GCG more straightforward, one can
parameterize them  as\footnote{In standard Calabi-Yau compactifications, one takes $\theta^1=\theta^2=\theta$, $\tilde \theta^i=0$. Note that in this case $\eta_i$ as defined in (\ref{spinorsIIA}) are not Majorana, but the combinations $\eta_1+\eta_2$ and $\iu(\eta_1-\eta_2)$ are.}
\begin{equation} \label{spinorsIIA}
\eta_1=\left(\begin{array}{c}
\theta^{1}_{+}\\
\tilde \theta^1_-
\end{array}\right), \qquad
\eta_2=\left(\begin{array}{c} \tilde \theta^2_+ \\
\theta^2_{-}
\end{array}\right) \ .
\end{equation}
For the special ansatz $\tilde \theta^i=0$, $L$ and $K$ have particularly nice forms in  in terms of the $O(6,6)$ pure spinors\footnote{\label{foot:JO}Note that for this ansatz, $J$ ($\Omega$) contains the bilinears between the two $\theta$'s involving two (three) Spin(6) gamma matrices.}
\beq
\Phi^\pm=e^B (\theta^1_+ \otimes \bar \theta^2_{\pm})
\eeq
the $SL(2,\mathbb{R})$ vielbein $u^i$ related to the four-dimensional axiodilaton, $i=1,2$, and the RR-spinor $C=C_1+C_3+C_5$ as \cite{Grana:2009im}
\begin{equation}
 L = \e^C (0, \Phi^+) \ , \qquad K_3+ \iu K_1 = \e^C (0,0, u^i \Phi^-) \ ,
\end{equation}
where this is written in the  $\Tsub$ decomposition of $\Es7$, given explicitly below in (\ref{56Tsub}), (\ref{133Tsub}), and $K_2$ is just the commutator of $K_3$ and $K_1$.


\section{From $N=2$ to $N=1$: Orientifolding and orbifolding in EGG}

\label{sec:Orbifolding}
As we will show in detail in Section \ref{sec:Kahlersub}, orientifolds break the $E_{7(7)}$ covariance into that of the subgroup $\Tsub_O$, where the ``$O$'' makes it explicit that this is a different subgroup from that of $T$ and $S$-duality.  Furthermore, as we will see, each orientifold projection gives rise to a different $\Tsub$ subgroup. The splitting of the fundamental and adjoint representations of $E_{7(7)}$ has been given in \eqref{56Tsub} and \eqref{133Tsub}.

\subsection{Orbifold action on $T_8$ and its reduction to M-theory}
\label{sec:OrbifoldingM}
The M-theory uplift of type IIA O6-orientifolds are a geometric involution on the 7-dimensional space. Such an involution can in turn be uplifted to an orbifold action $\tilde \sigma^*$ on the 8-dimensional space $T_8$ such that the SL(8) bundle decomposes at the locus of the action into a positive and a negative eigenbundle
\begin{equation} \label{decomp_V_orbifolding}
 T_8 = T_8^+ \oplus T_8^- \ ,
\end{equation}
and such that both subspaces are four-dimensional. The adjoint of $E_{7(7)}$ (\ref{decomposition_sl8_ad})
 decomposes under \eqref{decomp_V_orbifolding} as
\begin{equation}\begin{aligned}
E= & (T_8^+\otimes T_8^{*+})_0 \oplus (T_8^-\otimes T_8^{*-})_0\oplus (\Lambda^2 T_8^{*+} \otimes  \Lambda^2 T_8^{*-})\oplus T_0 \oplus \Lambda^4 T_8^{*+}  \oplus \Lambda^4 T_8^{*-} \\
& \oplus (T_8^+\otimes T_8^{*-}) \oplus (T_8^-\otimes T_8^{*+}) \oplus (\Lambda^3 T_8^{*+} \otimes  \Lambda^1 T_8^{*-})\oplus (\Lambda^1 T_8^{*+} \otimes \Lambda^3 T_8^{*-}) \ ,
\end{aligned} \end{equation}
where $T_0$ is the element of the adjoint that acts as $\pm 1$ on $T_8^\pm$. By comparing to \eqref{133Tsub}, we can see how $\E7$ is broken into $\Tsub$: the first line builds up the adjoint of $\Tsub_O$, which is even under the orbifold action, while the terms in the second line are odd and form the $({\bf 2},{\bf 32'})$ representation.
Similarly, the fundamental of $E_{7(7)}$, given in (\ref{decomposition_sl8_fund}) in terms of $SL(8)$ representations, decomposes as
\begin{equation}
A= \Lambda^2 T_8^+ \oplus \Lambda^2 T_8^- \oplus \Lambda^2 T_8^{*+} \oplus \Lambda^2 T_8^{*-} \oplus (T_8^+\otimes T_8^-) \oplus (T_8^{*-} \otimes T_8^{*-}) \ ,
\end{equation}
where the first four terms are even and form the $({\bf 12},{\bf 2})$ representation of $\TsubO$ (see Eq. (\ref{56Tsub})), while the last two, odd terms, form the $({\bf 32},{\bf 1})$.

To descend to M-theory, we require the orbifold action to have positive eigenvalue when acting on $\hat \rho_8$. Eq. (\ref{decomposition_gl7_V}) tells us then that $T_7^+$ is three-dimensional, while $T_7^-$ is four-dimensional. Finally, to recover the type IIA orbifold action that gives rise to O6 planes, we require $\hat v$ to have negative eigenvalue. In summary
\begin{equation} \label{T8O6}
T_8=T_8^+\oplus T_8^-=(T_7^+ \oplus T_{\hat \rho_8}^+) \oplus T_7^- = (T_6^+ \oplus T_{\hat \rho_8}^+) \oplus (T_6^- \oplus T_y^-) \ .
\end{equation}
We will come back to the full orientifold projection later in Section \ref{sec:O6}

Now let us see how the orbifold acts on the $N=2$ structures defined in the previous sections.
In type IIA, an involutive symmetry $\sigma$ that can be used to mod out the theory should be anti-holomorphic if $N=1$ supersymmetry is to be preserved
 \cite{Acharya:2002ag}. This means that for an $SU(3)$ structure defined by $J$ and $\Omega$, it should act as
$ \sigma^* J = - J$,  $\sigma^* \Omega =  \bar \Omega$.
This is easy to uplift to an action on $J_8$, $\Omega_4$ defining the SU(4) structure on $T_8$, namely  we require $\sigma$ to act as
\begin{equation}\label{sigma_J8_Omega4}
 \tilde \sigma^* J_8 = - J_8 \ , \qquad  \tilde \sigma^* \Omega_4 =  - \bar \Omega_4 \ .
\end{equation}
This implies that the action induced by $\tilde \sigma^*$ on $K_a$ and $L$ in the ${\bf 133}$ and the ${\bf 56}$ representation, which define an $SU(6)\subset \E7$ structure, should be
\begin{equation} \label{tildesigmaLK}
 \tilde \sigma^* L = - L \ , \qquad  \tilde \sigma^* K_1 = K_1 \ , \qquad \tilde \sigma^* K_{2/3} = - K_{2/3} \ .
\end{equation}
Therefore, $L$ and $K_{2/3}$ are not well-defined any more in the presence of fixed points. However, their products are well-defined under the orbifold action. More precisely, their product in the ${\bf 912}$ representation is defined by\footnote{Note that the ${\bf 56}$ component of their tensor product is forced to be zero by the compatibility condition.}
\begin{equation} \label{phi_from_KL}
 L \otimes (K_2+\iu K_3)  = \sqrt{\kappa} \phi  \ ,
\end{equation}
where the pre-factor $\sqrt{\kappa}$ appears due to the different normalizations in \eqref{normspinSU6} and \eqref{normspinSU7}.
This $\phi$ in the ${\bf 912}$ defines in turn an $SU(7)$ structure following Section~\ref{sec:SU(7)}, corresponding to the single spinor that survives the orbifolding. $\phi$ defines  the metric $g$ and a four-form $\phi_4$, cf.\ \eqref{phi_SU7}. From \eqref{phi_from_KL} we find
\begin{equation} \label{relationPhi4J8Omega8}\begin{aligned}
 \phi_4 = & \sqrt{\kappa}^{-1} \e^{-K_L/2} \Im \Omega_4 + \sqrt{\kappa} \e^{K_L/2} J_8 \wedge J_8  \, \\
 \phi^r_4 = & \e^{-K_L} \Im \Omega_4 + \kappa J_8 \wedge J_8 \ .
\end{aligned}\end{equation}
We find the second expression using $c = \sqrt{\kappa} \e^{-K_L/2}$.

The orbifold projection selects an $N=1$ special K\"ahler subspace inside the $N=2$ K\"ahler and quaternionic spaces.
The generator of its complex structure is given in \eqref{cs_SU7}, which can in turn be written in terms of the $SU(4)$-structure objects $J_8$ and $\Omega_4$ in $L$ and $K$ respectively as
\begin{equation}
 {\cal J}_\phi = \e^{A_4} \cdot (0,\  \e^{-K_L} \Im \Omega_4 + \kappa J_8 \wedge J_8) = \e^{-K_L} K_1 + \kappa {\cal J}_L \ .
\end{equation}
Therefore, the complex structure on the $N=1$ K\"ahler space $\e^{\cal J_\phi}$ is the tensor product of the complex structures $\e^{K_1}$ with $\e^{{\cal J}_L}$. If the orbifold singularities were blown-up (see comments below),  $\e^{\cal J_\phi}$ would not be block-diagonal any more.
As long as the singularities are not blown up, the K\"ahler potential \eqref{Kahlerpot_SU7} simplifies to the sum the two K\"ahler potentials for $L$ and $K$, i.e.\
\begin{equation}
 K_\phi = -\tfrac12 \log \kappa + K_L +\tfrac12 \log 2 \ ,
\end{equation}
The superpotential is given in \eqref{superpotential_SU7}. On the other hand, we have the Killing prepotentials \eqref{prepotentials_sol} that should descend to the $N=1$ description. Comparing both expressions we find
the relation
\begin{equation}
\e^{K_\phi/2} W = \tfrac12 \kappa^{-1} (P^2 + \iu P^3 ) \ .
\end{equation}

The above formulas are valid for the orbifold of an $SU(6)$ structure. If we blow-up the singularities resulting from the orbifolding, we switch on additional modes in $\phi$ that alter its form from the one given in \eqref{phi_from_KL}. More precisely, the objects $K_{2/3}$ and $L$ are not well-defined on the blown-up manifold, while $\phi$ still defines the geometry. The blow-up should lead to new modes that enter $\phi_4$ as extra four-forms.

\subsection{Further descent to type IIA} \label{sec:IIAorientifolds}

\subsubsection{O6 orientifolds} \label{sec:O6}

To recover O6-orientifolds, the orbifold involution $\tilde \sigma$ should have negative eigenvalue on $T_y$, i.e. act on $T_7$ as $\textrm{diag}(\sigma, -1)$ where $\sigma$ is an involution on the 6-dimensional space whose action is given above (\ref{sigma_J8_Omega4}).
Furthermore, the O6 projection mods out by the action of $\sigma \Omega_p (-1)^{F_L}$, where $\Omega_p$ is the worldsheet parity, and $(-1)^{F_L}$ gives an additional minus sign on the RR sector. The uplift of this combination is the purely geometric involution $\tilde \sigma$.

The combined operation $\Omega_p (-1)^{F_L}$ has a different action on the different $\Tsub$ components of $L$ and $K_a$. On $O(6,6)$ bispinors, such as the RR potentials or the pure spinors of generalized complex geometry $\Phi^\pm$, which are tensor products of a left and a right-moving spinor, it acts in the following way \cite{Benmachiche:2006df}
\beq
(-1)^{F_L} \Omega_p  \Phi^+ = \lambda(\Phi^+) \ , \qquad (-1)^{F_L} \Omega_p  \Phi^- = - \lambda(\bar \Phi^-) \ , \qquad (-1)^{F_L} \Omega_p  C =  \lambda(C)
\eeq
where $\lambda$ is the following action on forms
\begin{equation} \label{lambda}
 \lambda (\alpha_{2p}) = (-1)^p \alpha_{2p} \ , \qquad \lambda (\alpha_{2p-1}) = (-1)^p \alpha_{2p-1} \ .
\end{equation}
This can be understood since worldsheet parity exchanges the left and right-moving sectors, and on the bispinors, which are tensor products of left and right moving spinors, it acts by transposition.
Since the orientifold projection keeps states which are even under the action of $\sigma (-1)^{F_L} \Omega_p$, one requires the involution to satisfy
\beq
\sigma^*  \Phi^+ = \lambda(\Phi^+) \ , \qquad \sigma^*  \Phi^- = -\lambda(\bar \Phi^-) \ , \qquad \sigma^*  C =  \lambda(C) \ .
\eeq
We want to define an analogous ``$\lambda$-operation" as an action on fundamental $SL(2)$ and $O(6,6)$ indices. The following operator acting respectively on the ${\bf 12}$ of $O(6,6)$, ${\bf 2}$ of $SL(2,\mathbb{R})$ and ${\bf 32}$ of $O(6,6)$ does the job
\begin{equation}
 \tilde \lambda = \left(\begin{aligned} 1_{6\times 6} && 0 \\ 0 && -1_{6\times 6} \end{aligned}\right)
\otimes \left(\begin{aligned} 1 && 0 \\ 0 && -1 \end{aligned}\right) \otimes (-\lambda)  .
\end{equation}
 For higher representations $\tilde \lambda$ just acts on all indices.\footnote{Note that the action on $T\oplus T^*$ then induces the action $\lambda$ in (\ref{lambda}) on the spinor representation (isomorphic to a sum of forms).}.
Therefore, on the fundamental ${\bf 56}$ representation, which decomposes into $\Tsub$ as in (\ref{56Tsub}), this action reads
\begin{equation} \label{lambdaL}
  \tilde \lambda (L) = \left( \left(\begin{aligned} L^{\hm1} && - L_{\hm}{}^{1} \\  - L^{\hm 2} && L_{\hm}{}^2 \end{aligned}\right), -\lambda(L^+) \right) \ .
\end{equation}
On the adjoint representation, whose $\Tsub$ decomposition is given in (\ref{133Tsub}), we get
\begin{equation} \label{lambdaK}
  \tilde \lambda (K) = \left( \left(\begin{aligned} K^{\hm}{}_{\hn} && - K^{\hm\hn} \\ - K_{\hm\hn} && K_{\hm}{}^{\hn} \end{aligned}\right), \left(\begin{aligned} K^{1}{}_{1} && - K^{1}{}_2 \\ - K^2{}_1 && K^2{}_2 \end{aligned}\right),\left(\begin{aligned} -\lambda(K^{-1}) \\ \lambda(K^{-2}) \end{aligned}\right)\right) \
\end{equation}
(where by construction $K^2{}_2=-K^1{}_1$ and $K_{\hm}{}^{\hn}=-K^{\hm}{}_{\hn}$).

The claim is that $\tilde \lambda $ acts like $\Omega_p (-1)^{F_L}$ on $K_1$, while on $L$ and $K_{2/3}$ it is $-\tilde \lambda$ that does the job. The fields that will survive the orientifold projection are therefore those for which $\sigma$ acts in the following way
\begin{equation}\label{sigmaLK}
\sigma^*L=- \tilde \lambda(L) \ , \qquad \sigma^* K_1 = \tilde \lambda(K_1) \ , \qquad \sigma^* K_{2/3} = - \tilde \lambda(K_{2/3}) \ .
\end{equation}
In the language of \eqref{spinorsIIA}, we see from \eqref{L0}-\eqref{K0} that
the action of $\tilde \lambda$ corresponds to the exchange of the two spinors $\eta^i$.

For the corresponding vector fields $L_\mu= (L_{\mu}^{iA}, L_\mu^+)$ in the ${\bf 56}$ representation we have
\begin{equation} \label{sigmaLmu}
  \sigma^* L_\mu = \tilde \lambda (L_\mu) \ .
\end{equation}
Here, $L_{\mu}^{Ai}$ are the electric and dual magnetic vectors coming from the off-diagonal components fo the metric and the B-field, while $L_\mu^+$ collects the Ramond-Ramond fields with one external leg, i.e.\ $L_\mu^+ = (C_1)_\mu + (C_3)_\mu + (C_5)_\mu + (C_7)_\mu$.

\subsubsection{Type IIB orientifolds}

For completeness (and because we will use later O9 as an illustration) we give the action for type IIB orientifolds.\footnote{Note that in type IIB the roles of $\Phi^+$ and $\Phi^-$ are exchanged.} There, the theory is modded out by $\sigma \Omega_p (-1)^{F_L}$ for O3/O7 projection, and $\sigma \Omega_p$ for O5/O9. This means that the latter projection has an extra minus sign on the $O(6,6)$ spinors with respect to the type IIA case, i.e. we define
\begin{equation} \label{lambdaIIB}
 \tilde \lambda_{\rm{IIB}} = \left(\begin{aligned} 1_{6\times 6} && 0 \\ 0 && -1_{6\times 6} \end{aligned}\right)
\otimes \left(\begin{aligned} 1 && 0 \\ 0 && -1 \end{aligned}\right) \otimes \pm \lambda  ,
\end{equation}
where the plus sign is for O3/O7 projections, while the minus applies to O5/O9. This contributes to a $\pm$ sign in the last components of (\ref{lambdaL}) and (\ref{lambdaK}). We then require (\ref{sigmaLK}) and \eqref{sigmaLmu}, with $\tilde \lambda$ replaced by $\tilde \lambda_{\rm{IIB}}$.

\subsubsection{New ${\mathbb Z}_2$ projections}
In general, new orientifold actions can be found by conjugating known orientifold actions with elements in $E_{7(7)}(\mathbb{Z})$. For all these new orientifolds our discussion applies. A simple example of a new $\mathbb{Z}_2$ action is the NS5-projection that is related to O5-orientifolding in type IIB by S-duality in ten dimensions. Concerning the involution, S-duality only exchanges the roles of the $B$ field and the field $C_2$. Therefore, the NS-NS (R-R) sector is even (odd) under the resulting involution, which thus can be written as $(-1)^{F_L} \sigma$, where $F_L$ is the left-moving fermion number on the world-sheet and $\sigma$ is an involution of the internal space that inverts four of the internal directions, satisfying the following on the pure spinors of GCG and the RR fields
\begin{equation}
\sigma^* \Phi^\pm = \Phi^\pm \ , \qquad \sigma^* C = - C \ .
\end{equation}
Correspondingly, the involution $\tilde \lambda_{\rm NS5}$ is
\begin{equation}
\tilde \lambda_{\rm NS5} = 1_{12\times 12}\otimes 1_{2\times 2} \otimes (-1)_{\bf 32} \ .
\end{equation}
Subsequently, the roles of $K_1$ and $K_3$ are exchanged with respect to \eqref{tildesigmaLK}. S-duality implies that the fixed points of this action are negative tension objects with negative NS5-brane charge.

Note that this action can be defined in a completely analogous way in type IIA, and it can be uplifted to M-theory to find the orientifolding for M5-branes, with
\begin{equation}
\tilde \sigma^*_{\rm M5} \diff y = -\diff y \ , \qquad \tilde \sigma^*_{\rm M5} \rho_8 = - \rho_8 \ .
\end{equation}
The analogous D4-orientifolding lifts to the same expression, but without any involution of $\diff y$. Though we can uplift these involutions to M-theory, even with these assignments for $\tilde \sigma_{\rm M5}$, the M5-involution does not become an orbifold action. More precisely, it acts on the adjoint representation as $\sigma^*_{\rm M5}$ but on the exceptional generalized tangent bundle with an extra minus sign, i.e.\ with $-\sigma^*_{\rm M5}$. Therefore, the projection on $L$ and $K_a$ is given by
\begin{equation} \label{tildesigmaLKM5}
 \tilde \sigma^*_{\rm M5} L = L \ , \qquad  \tilde \sigma^*_{\rm M5} K_1 = K_1 \ , \qquad \tilde \sigma^*_{\rm M5} K_{2/3} = - K_{2/3} \ ,
\end{equation}
in contrast to the orbifold action given in \eqref{tildesigmaLK}. Thus, no $SU(7)$ structure surviving the involution can be defined.
As a consequence the fixed points of $\sigma_{\rm M5}$ cannot be resolved in a geometric way within M-theory, in contrast to the orbifold fixed points of Section~\ref{sec:OrbifoldingM}.

\subsubsection{K\"ahler subspaces} \label{sec:Kahlersub}

Here we show how the orientifold projection selects the $N=1$ special K\"ahler subspaces inside the $N=2$ K\"ahler and quaternionic ones.
Before we analyze how a K\"ahler space emerges from the projection on the hypermultiplets, we first want to understand the reduction of $E_{7(7)}$ under the orientifolding. Let us first start with an orientifolding to O9-planes, i.e.\ $\sigma$ is the identity (and we are modding out the theory just by the action of $\Omega_p$).
As the representations of $E_{7(7)}$ split into the even and odd parts under $\tilde \lambda$, they form representations of a subgroup of $E_{7(7)}$ that is the subgroup of even transformations. Therefore, we analyze the action of $\tilde \lambda$ on the adjoint of $E_{7(7)}$, split into representations of the subgroup  $O(6,6)\times SL(2,\mathbb{R})$ corresponding to T- and S-duality subgroups, cf.\ Eq. (\ref{133Tsub}).\footnote{\label{foot:TS}Here S-duality does not refer to the type IIB S-duality acting on  the RR axion $C_0$ and the dilaton, but the S-duality within the NS sector, that acts on $\tau=B_6+ \iu e^{-\phi}$.}

On the adjoint of $O(6,6)$, $\tilde \lambda$ acts as in the first component of (\ref{lambdaK}). The ${\bf 66}$ representation therefore gets projected to
\begin{eqnarray} \label{Oproj66}
 {\bf 66}  &\to&  {\bf 35}_{\bf 0} \oplus {\bf 1}_{\bf 0}\ , \\
 \mu^A{}_B &\to& \mu^{\hm}{}_{\hn} \nn
\end{eqnarray}
Thus we find $O(6,6)\to Gl(6,\mathbb{R}) = SL(6,\mathbb{R}) \times \mathbb{R}_+$ (and the subindex denotes the charge under this $\mathbb{R}_+$, corresponding to the volume). Similarly
\begin{eqnarray} \label{Oproj3}
 {\bf 3} & \to&  {\bf 1}_{\bf 0} \ , \\
 \mu^i{}_j & \to& \mu^1{}_1 \nn
\end{eqnarray}
and thus $SL(2,\mathbb{R}) \to \mathbb{R}_+$ (and the subindex denotes the charge under this $\mathbb{R}$, corresponding to the dilaton). Finally, on the $({\bf 32'},{\bf 2})$ representation we get (recall that for O9, $\tilde \lambda_{\rm{IIB}}$ acts as $-\lambda$ on the spinor)
\begin{eqnarray} \label{Oprojection_spinor}
 ({\bf 32'},{\bf 2}) &\to& {\bf 1}_{({\bf +1},{\bf +1})} \oplus{\bf 15}_{({\bf +1},{\bf -1})} \oplus {\bf 1}_{({\bf -1},{\bf -1})} \oplus {\bf 15'}_{({\bf -1},{\bf +1})} \ , \\
 \mu^{+i} & \to& \mu_{(0)}^2+\mu_{(2)}^1+\mu_{(6)}^1+\mu_{(4)}^2 \  \nn .
\end{eqnarray}
On the first line we stated the $SL(6,\mathbb{R})$ representations and denoted the charges under the two $\mathbb{R}_+$ factors coming from the volume and the dilaton, and on the second line the superscript denotes the $SL(2)$ component, while the number in parenthesis in the subscript denotes the degree of the form.

We see that the diagonal $\mathbb{R}_+$ factor together with the two scalars in (\ref{Oprojection_spinor}) forms an $SL(2,\mathbb{R})$ group, while the two ${\bf 15}$s together with the non-diagonal $\mathbb{R}_+$ factor enhance the $SL(6,\mathbb{R})$ to $O(6,6)$. Thus, the new covariance group is $O(6,6)\times SL(2,\mathbb{R})$. Since this is a different $\Tsub$ from the original one associated with S- and T-duality (see Footnote~\ref{foot:TS}), we call this $\TsubO$. Thus we get
\beq
\begin{aligned}
E_7 \to & \Tsub_{O_9} \ , \\
\mu \to & \left( \left(\begin{aligned} \mu^{\hm}{}_{\hn} && \mu^{1}_{(2)} \\ \mu^{2}_{(4)} && \mu_{\hm}{}^{\hn} \end{aligned}\right), \left(\begin{aligned} \mu^{1}{}_{1} && \mu^{1}_{(6)} \\ \mu^{2}_{(0)} && -\mu^{1}{}_{1} \end{aligned}\right), 0\right) \ , \nn \\
A_{\rm{shifts}}\to& \left( \left(\begin{array}{cc} 0 & C_2 \\ 0 & 0 \end{array}\right), \left(\begin{array}{cc} 0 & C_6 \\ 0 &0 \end{array}\right), 0 \right)
\end{aligned} \ .
\eeq
where in the last line we have used the $\Tsub$ embedding of the $B$ and $C$-fields given in (\ref{AshiftsTsub}).


If $\sigma$ is not the identity, the situation is slightly more involved. The orientifolding in general maps different points onto each other. Only at the locus of the O-planes the covariance group can be really projected to a subgroup. Let us consider the case of an O$(3+p)$-plane. At the O$(3+p)$-plane we can split the tangent space
\beq
T \to T^{(p)}_\parallel \oplus T^{(6-p)}_\perp \ ,
\eeq
 where the supraindex in parenthesis indicates the dimensions of each space. The involution $\sigma^*$ acts as $+1$ on $T_\parallel$ and as $-1$ on $T_\perp$. Therefore, the combination $\sigma^* \tilde \lambda$ projects the geometric group
 \beq
 Gl(6,\mathbb{R})\to Gl_\parallel(p,\mathbb{R})\times Gl_\perp(6-p,\mathbb{R}) \ .
 \eeq
  Furthermore, $\sigma^* \tilde \lambda$ projects
\beq
  \Lambda^2 T^* \to T^*_\parallel \otimes T^*_\perp \ ,  \qquad \Lambda^2 T \to T_\parallel \otimes T_\perp \ ,
\eeq
and these give each the $({\bf p},{\bf 6-p})$ representations that enhance $Gl_\parallel(p,\mathbb{R})\times Gl_\perp(6-p,\mathbb{R})$ to $Gl(6,\mathbb{R})_{O(3+p)}$. Thus, as for the case of O9-planes, we find $O(6,6)\to Gl(6,\mathbb{R})$, but now to a different $Gl(6,\mathbb{R})$ indicated by the subindex.
Under the breaking $O(6,6)\to Gl(6,\mathbb{R})_{O_{(3+p)}}$, the $({\bf 2},{\bf 32'})$ representation projects as in \eqref{Oprojection_spinor}. Here, the two surviving singlets are $\Lambda^p T^*_\parallel$ and $\Lambda^{(6-p)} T^*_\perp$, which form singlets under the emerging $Gl(6,\mathbb{R})_{O(3+p)}$. Hence, we see that for all orientifold actions, the covariance group projects to $\TsubOp$ (as we saw this subgroup is different for each type of orientifold).

Now let us consider the projection \eqref{sigmaLK} on the vector and hypermultiplet sectors. The vector fields that survive the orientifold projection are those that are even under $\sigma^* \tilde \lambda$ (see (\ref{sigmaLmu})).  For O9, where $\sigma$ is the identity and $\tilde \lambda$ acts as $-\lambda$ in the spinor part, we get that the surviving vector fields are the Kaluza-Klein vectors as well as the vectors associated with  the internal one-form $(C_2)_{\mu}$ (and their magnetic duals).
As for the $N=1$ chiral fields that descend from $N=2$ vector multiplets, we keep from $L$ only the pieces that are invariant under $-\sigma^* \tilde \lambda$.  Again, for an O9 and an SU(3) structure, we have that the projection onto states that are invariant under $\tilde \lambda$ gives that all degrees of freedom in the three-form $\Omega$ is kept.

In the hypermultiplet sector, since $E_{7(7)}$ is projected onto $\TsubOp$, we know from \eqref{sigmaLK} that both $K_2$ and $K_3$ will be in the $({\bf 32'},{\bf 2})$ representation of $\TsubOp$, which is odd under the projection. Since $[K_2,K_3] \sim K_1$ and each $K_2$ and $K_3$ determine each other, $K_2 + \iu K_3$ defines a pure $O(6,6)$ spinor tensored with a doublet of $SL(2,\mathbb{R})$.
Therefore, $K_2 + \iu K_3$ parametrizes a special K\"ahler space. On the other hand, $K_1$ is in the $({\bf 66},{\bf 1})\oplus ({\bf 1},{\bf 3})$ of $\TsubOp$. It can be understood as a generalized almost complex structure on the orbit of $K_2 + \iu K_3$.

\section{Discussion}\label{section:Conclude}
In this work we derived the form of the couplings for general $SU(7)$ and $SU(6)$ structures in M-theory and type IIA, which correspond to (off-shell) $N=1$ and $N=2$ supersymmetric compactifications to four dimensions, building on and extending the work of \cite{Pacheco:2008ps,Grana:2009im}.
Using EGG we could reformulate all degrees of freedom in such backgrounds by a set of fundamental objects in $E_{7(7)}$ representations. Moreover, the effective couplings are easily determined as singlets that are tensor products of the fundamental objects and their first derivatives in $E_{7(7)}$. In particular, $N=1$ backgrounds are determined by an $SU(7)$ structure $\phi$ in the ${\bf 912}$ representation. Its quartic invariant gives the K\"ahler potential, while the superpotential is determined by an eigenvalue equation. In contrast, $N=2$ backgrounds admit two sectors: vector- and hyper-multiplets. The former is described by one object $L$ in the (fundamental) ${\bf 56}$ representation whose quartic invariant gives the K\"ahler potential. The hypermultiplets are described by an $SU(2)$ subalgebra spanned by a triplet of structures  $K_a$ in the adjoint representation.  The normalization of the $SU(2)$ commutator relations gives the hyper-K\"ahler potential of the hyper-K\"ahler cone over this quaternionic K\"ahler space. $L$ and $K_a$ together define an $SU(6)$ structure. The couplings of the two sectors, i.e. the prepotentials, are given by a triple tensor product of these two objects with the derivative operator.

Furthermore, we discussed involutions in EGG that are supposed to project an $N=2$ background to an $N=1$ one. Examples of these involutions are orbifoldings in M-theory or orientifoldings in type II. We found the explicit map between the original $SU(6)$ structure and its $SU(7)$ descendant. In particular, while $L$ as well as $K_2+ \iu K_3$ are projected out, their tensor product produces $\phi$, which defines the $SU(7)$ structure and survives the blow-up to a smooth geometry. The $N=1$ K\"ahler potential and superpotential are then naturally determined by the $N=2$ K\"ahler and hyper-K\"ahler potentials, and the prepotentials. We also determined the projection that creates negative tension objects with negative M5-brane charge in M-theory and observed that no $SU(7)$ structure can be defined in that case, i.e.\ one cannot describe the resolution of singularities from involutions other than orbifoldings in EGG. In particular, the orientifold singularities related to D6-branes in type IIA cannot be resolved, but the corresponding M-theory orbifold fixed points can. In other words, the pure existence of an extra coordinate enables to resolve the singularities of D6-branes and O6-planes. It seems in order to describe D-branes in EGG one needs to introduce extra coordinates. For instance, while NS5-branes cannot be described in generalized geometry, they could in principle be described in doubled geometry. It would be interesting to understand the resulting doubled geometries in the presence of NS5-branes and their negative tension counterparts further. Even more challenging would be the realization of a 56-dimensional space that covariantizes $E_{7(7)}$ and could describe all branes in type II string theory or M-theory.

We pointed moreover out that there exists an intermediate generalized tangent bundle $T_8$ in M-theory that transforms under an $SL(8)$ subgroup of $E_{7(7)}$. From the type IIA point of view, this $SL(8)$ group contains the geometric transformations $SL(6)$ and the group $SL(2)$ transforming the four-dimensional axiodilaton $\tau = B_6 + \iu \e^{-\phi}$.
In this language, $N=1$ backgrounds are described by a four-form on $T_8$, $N=2$ backgrounds by a real two-form and a complex four-form, i.e.\ they correspond to $Spin(7)$ and $SU(4)$ structures in eight dimensions. This suggests that there should exist a lift to an eight-dimensional space $M_8$ on which $T_8$ is the tangent bundle, similar to F-theory, whose volume is normalized everywhere. In the fashion described above, $M_8$  would not only geometrize D6-branes, but also some kind of exotic branes (as described in \cite{Bergshoeff,deBoer:2012ma}) that form a set of $(p,q)$-branes for the four-dimensional axiodilaton, similar to F-theory. It would be very interesting to understand such geometries further.

\section*{Acknowledgments}
We would like to thank Diego Marqu\'es, Ruben Minasian, Eran Palti and Daniel Waldram for useful discussions. This work was supported in part by the ANR grant 08-JCJC-0001-0 and the ERC Starting Grants 259133 -- ObservableString and 240210 - String-QCD-BH.

\vskip 1.5cm

\appendix

\noindent
{\bf\Large Appendix}

\section{$E_{7(7)}$ group theory} \label{App:E7}
In the following we give relevant formulas for products of representations in the $SL(8)$ decomposition and their translation into the $SU(8)$ decomposition of spinors, following and extending \cite{Grana:2011nb}.

\subsection{$E_{7(7)}$ group theory in terms of $SL(8)$ representations}

The $E_{7(7)}$ representations of interest are the fundamental ${\bf 56}$, the adjoint ${\bf 133}$ and the ${\bf 912}$, whose decompositions under $SL(8)$ are given in \eqref{decomposition_sl8_fund}, \eqref{decomposition_sl8_ad} and \eqref{rep912}, respectively. In the following we denote objects in the fundamental representation by $\alpha$, $\beta$, etc.\ and write them in terms of $SL(8)$ representations as
\begin{equation}
 \alpha = ( \alpha^{ab} ,\, \alpha_{ab}) \ .
\end{equation}
Similarly, the adjoint representation decomposes as
\begin{equation}
 \mu = ( \mu^a{}_b ,\, \mu_{abcd}) \ .
\end{equation}
The ${\bf 912}$ representation finally is given by
\begin{equation}
 \phi = (\phi^{ab},\, \phi^{abc}{}_d,\, \phi_{ab},\, \phi_{abc}{}^d) \ .
\end{equation}

We will use the following notation for the product of representations:
\begin{align} \label{tensorprodnotation}
( , ) &: [{\rm{\bf rep}} \otimes {\rm{\bf rep}}]_ {\bf 1} \nn \ , \\
\times &: [{\rm{\bf rep}} \otimes {\rm{\bf rep}}]_ {\rm{\bf 133}} \nn \ , \\
\cdot &: [{\bf 133} \otimes {\rm{\bf rep}}]_ {\rm{\bf rep}}  \ , \\
\otimes &: [{\rm{\bf 56}} \otimes {\rm{\bf 133}}]_ {\rm{\bf 912}} \nn \ , \\
\otimesF &: [{\rm{\bf 912}} \otimes {\rm{\bf 133}}]_ {\rm{\bf 56}} \nn \ , \\
\otimesA &: [{\rm{\bf 912}} \otimes {\rm{\bf 56}}]_ {\rm{\bf 133}} \nn \ ,
\end{align}
where {\bf rep} is any representation of $\Es7$.

The action of the adjoint on the fundamental representation, in other words the product ${\bf 133} \times {\bf 56} \to {\bf 56}$, is given by
\begin{equation}\begin{aligned}
(\mu \cdot \alpha)^{ab} = & \mu^a{}_c \alpha^{cb} +  \mu^b{}_c \alpha^{ac} + (\vol^{-1}_8 \llcorner  \mu)^{abcd} \alpha_{cd} \ , \\
(\mu \cdot \alpha)_{ab} = & - \mu^c{}_a \alpha_{cb} - \mu^c{}_b \alpha_{ac} - \mu_{abcd} \alpha^{cd} \ .
\end{aligned}\end{equation}
The symplectic invariant on the ${\bf 56}$ reads
\begin{equation} \label{symplinvt}
 \langle \alpha , \beta \rangle = \alpha^{ab} \beta_{ab} - \alpha_{ab} \beta^{ab} \ .
\end{equation}
The trace in the adjoint is
\begin{equation} \label{tradj}
(\mu, \nu)= \mu^a{}_b \nu^b{}_a - \frac16 (\vol^{-1}_8 \llcorner \mu)^{abcd} \nu_{abcd} \ .
\end{equation}

The ${\bf 56} \times {\bf 56} \to {\bf 133}$ reads
\begin{equation}\begin{aligned} \label{56x56=133}
 (\alpha \times \beta)^a{}_b = &  \alpha^{ca} \beta_{cb} - \tfrac18 \delta^a_b \alpha^{cd} \beta_{cd} + \alpha_{cb} \beta^{ca} - \tfrac18 \delta^a_b \alpha_{cd} \beta^{cd}  \ , \\
 (\alpha \times \beta)_{abcd} = & - 3 (\alpha_{[ab} \beta_{cd]} + \tfrac{1}{4!} \epsilon_{abcdefgh} \alpha^{ef} \beta^{gh} ) \ .
\end{aligned}\end{equation}
The action of the adjoint onto itself, i.e.\ the ${\bf 133} \times {\bf 133} \to {\bf 133}$ is given by
\begin{equation}\begin{aligned}
 (\mu \cdot \nu)^a{}_b = & (\mu^a{}_c \nu^c{}_b -\mu^c{}_b \nu^a{}_c) - \tfrac13 ((\vol^{-1}_8 \llcorner  \mu)^{acde} \nu_{bcde} - \mu_{bcde} (\vol^{-1}_8 \llcorner \nu)^{acde})  \ , \\
 (\mu \cdot \nu)_{abcd} = & 4 (\mu^e{}_{[a} \nu_{bcd]e} - \nu^e{}_{[a}\mu_{bcd]e} )\ .
\end{aligned}\end{equation}
The product ${\bf 56} \times {\bf 133} \to {\bf 912}$ is given by
\begin{equation}\label{cross_product_sl8}\begin{aligned}
 (\alpha \otimes \mu)^{ab} = & \mu^a{}_c \alpha^{cb} + \mu^b{}_c \alpha^{ca} \ , \\
 (\alpha \otimes \mu)^{abc}{}_d = & -3(\alpha^{[ab} \mu^{c]}{}_d - \tfrac13 \alpha^{e[a} \mu^b_e \delta^{c]}_d ) +2 ((\vol^{-1}_8 \llcorner \mu)^{abce} \alpha_{ed} + \tfrac12  \alpha_{ef} ( \vol^{-1}_8 \llcorner \mu)^{ef[ab}\delta_d^{c]}) \ , \\
 (\alpha \otimes \mu)_{ab} = & -\mu^c{}_a \alpha_{cb} - \mu^c{}_b \alpha_{ca} \ ,\\
 (\alpha \otimes \mu)_{abc}{}^d = & -3(\alpha_{[ab} \mu_{c]}{}^d - \tfrac13 \alpha_{e[a} \mu_b{}^e \delta_{c]}^d ) +2 ( \mu_{abce} \alpha^{ed} + \tfrac12  \alpha^{ef}\mu_{ef[ab}\delta^d_{c]}) \ .
\end{aligned} \end{equation}
The adjoint action ${\bf 133} \times {\bf 912} \to {\bf 912}$ on the ${\bf 912}$ is given by
\begin{equation}\label{912adjoint_sl8}\begin{aligned}
 (\mu \cdot \phi)^{ab} = & \mu^a{}_c \phi^{cb} + \mu^b{}_c \phi^{ac} + \tfrac23 ((\vol^{-1}_8 \llcorner  \mu)^{cde(a} \phi^{b)}_{cde}) \ , \\
 (\mu \cdot \phi)^{abc}{}_d = & 3 \mu^{[a}{}_e \phi^{bc]e}{}_d - \mu^e_d \phi^{abc}{}_e + (\vol^{-1}_8 \llcorner  \mu)^{abce} \phi_{ed} + (\vol^{-1}_8 \llcorner \mu)^{ef[ab} \phi_{efd}^{c]} \\ &
 - ( \vol^{-1}_8 \llcorner \mu)^{efg[a} \phi_{efg}^b \delta^{c]}_d \ , \\
 (\mu \cdot \phi)_{ab} = &  - \mu^c_a \phi_{cb} - \mu^c{}_b \phi_{ac} + \tfrac23 ( \mu_{cde(a} \phi^{cde}{}_{b)}) \ ,\\
 (\mu \cdot \phi)_{abc}{}^d = & -3 \mu^e{}_{[a} \phi_{bc]e}{}^d + \mu^d{}_e \phi_{abc}{}^e + \mu_{abce} \phi^{ed} + \mu_{ef[ab} \phi^{efd}{}_{c]}- \mu_{efg[a} \phi^{efg}{}_b \delta_{c]}^d \ .
\end{aligned} \end{equation}
The product $ {\bf 912} \times {\bf 133}\to {\bf 56}$ reads
\begin{equation}\label{912x133_sl8}\begin{aligned}
 (\phi \otimesF \mu)^{ab} = & - (\phi^{ac} \mu^b{}_c - \phi^{bc} \mu^a{}_c) - 2 \phi^{abc}_d \mu^d{}_c \\
& + \tfrac23 (\phi_{cde}^a (\vol^{-1}_8 \llcorner \mu)^{cdeb} - \phi_{cde}^b (\vol^{-1}_8 \llcorner \mu)^{cdea}) \ , \\
 (\phi \otimesF \mu)_{ab} = &  (\phi_{ac} \mu^c{}_b - \phi_{bc} \mu_a^c) - 2 \phi_{abc}{}^d \mu^c{}_d \\
&  - \tfrac23 (\phi^{cde}{}_a \mu_{cdeb} - \phi^{cde}{}_b  \mu_{cdea}) \ .
\end{aligned} \end{equation}
Similarly, the product $ {\bf 912} \otimes {\bf 56}\to {\bf 133}$ reads
\begin{equation}\label{912x56_sl8}\begin{aligned}
 (\phi \otimesA \alpha)^{a}{}_{b} = & \alpha^{ca} \phi_{cb} + \alpha_{cb} \phi^{ca} + \alpha_{cd} \phi^{cda}{}_b -  \alpha^{cd} \phi_{cdb}{}^{a} \ , \\
 (\phi \otimesA \alpha)_{abcd} = &  -4 (\phi_{[abc}{}^e \alpha_{d]e} - \tfrac{1}{4!} \epsilon_{abcdefgh} \phi^{efg}{}_i \alpha^{hi}) \ .
\end{aligned} \end{equation}
The product ${\bf 912} \otimes {\bf 912} \to {\bf 133}$ reads in the $SL(8)$ decomposition
\begin{equation}\begin{aligned}
(\phi \times \psi)^a_b = & \phi^{ac} \psi_{cb} + \phi_{bc}\psi^{ca} + \tfrac13 (\phi_{cde}{}^a \psi^{cde}{}_b + \phi^{cde}{}_b \phi_{cde}{}^a)\\& - (\phi^{acd}{}_e\psi_{bcd}{}^e + \phi_{bcd}{}^e \psi^{acd}{}_e)\\
& - \tfrac18 \delta^a_b (\phi^{cd}\psi_{cd} + \phi_{cd}\psi^{cd} - \tfrac23 \phi^{cde}{}_f \psi_{cde}{}^f - \tfrac23 \phi_{cde}{}^f \psi^{cde}{}_f) \ , \\
(\phi \times \psi)_{abcd} = & 4 (\phi_{e[a} \psi_{bcd]}{}^e + \phi_{[bcd}{}^e\psi_{a]e} +\tfrac{1}{4!} \epsilon_{abcdefgh} (\phi^{ei} \psi^{fgh}{}_i + \phi^{fgh}{}_i \psi^{ei})) \\ & +2 (\phi_{f[ab}{}^e\psi^f_{cd]e} + \tfrac{1}{4!} \epsilon_{abcdefgh} \phi^{efj}{}_i \psi^{ghi}{}_j) \ .
\end{aligned} \end{equation}

\subsection{The relation to $SU(8)$ representations}
\label{sec:SU8grouptheory}
The $SU(8)$ representation is spanned by anti-symmetric products $\gamma_{ab}$ of the matrices $\gamma_a$ that obey the Clifford algebra
\begin{equation}
 \{\gamma_a, \gamma_b\}^\alpha{}_\beta = 2 g_{ab} \delta^\alpha_\beta \ .
\end{equation}
Furthermore, the gamma matrices $\gamma_a$ fulfill
\begin{equation}
 (\gamma^a)^\alpha{}_\beta (\gamma_a)^\gamma{}_\delta = \delta^\alpha_\delta \delta^\gamma_\beta \ .
\end{equation}
Under $SU(8)$, the $\rep{56}$ decomposes according to
\begin{equation}\label{suf} \begin{aligned}
\alpha&=(\alpha^{\alpha\beta},\bar{\alpha}_{\alpha\beta}) \ ,\\
\bf{56}&=\bf{28}\oplus \bar{\bf{28}} \ ,
\end{aligned}\end{equation}
while for the adjoint \textbf{133} we have
\begin{equation}\label{sua} \begin{aligned}
\mu&=(\mu^{\alpha}_{\,\,\,\beta},\mu^{\alpha\beta\gamma\delta},\bar{\mu}_{\alpha\beta\gamma\delta}) \ , \\
\bf{133}&=\bf{63}\oplus \bf{35}\oplus \bar{\bf{35}}\ .
\end{aligned}\end{equation}
where $\mu^\alpha{}_\alpha=0$ and $\bar{\mu}_{\alpha\beta\gamma\delta}=\ast_8 \mu_{\alpha\beta\gamma\delta}$.
Furthermore, we have for the $\rep{912}$ the $SU(8)$ decomposition
\begin{equation}\label{su912}\begin{aligned}
\phi&=(\phi^{\alpha\beta}, \phi^{\alpha \beta \gamma}{}_\delta, \bar{\alpha}_{\alpha\beta}, \bar \phi_{\alpha \beta \gamma}{}^\delta) \ ,\\
\bf{912}&=\bf{36}\oplus \bf{420}\oplus \bar{\bf{36}}\oplus \bar{\bf{420}} \ ,
\end{aligned}\end{equation}
Note that these are very similar to the $\SLE$ decompositions (\ref{decomposition_sl8_fund}), (\ref{decomposition_sl8_ad}) and \eqref{rep912}.  To go from one to the other, we use for the $\rep{56}$ \cite{Pacheco:2008ps}
\begin{equation}\label{fundsu8tosl8} \begin{aligned}
\alpha^{ab}&= (\alpha^{\alpha\beta}+ \bar \alpha^{\alpha\beta}) \gamma^{ab}{}_{\beta\alpha} \ , \\
\tilde \alpha_{ab}&=-\iu (\alpha^{\alpha\beta}- \bar \alpha^{\alpha\beta}) \gamma^{ab}{}_{\beta\alpha} \ ,
\end{aligned}\end{equation}
where we defined $\gamma^{ab}{}_{\beta\alpha}= C_{\beta \gamma} (\gamma^{ab})^\gamma{}_{\alpha}$ and $C_{\alpha \beta}$ is the matrix that induces transposition on spinors.
In the ${\bf{133}}$, if only the $\rep{63}$ adjoint representation of $SU(8)$ is nonzero,
i.e. if
 $\mu_{\alpha\beta\gamma\delta}=0$, one recovers the following $\SLE$ components
\begin{equation}\label{pw2} \begin{aligned}
\mu_{ab}&=\tfrac{\iu}{2} \mu^{\alpha}{}_{\beta} \gamma_{ab}{}^{\beta}{}_{\alpha} \ , \\
\mu_{abcd}&= \tfrac{1}{4} \mu^{\alpha}{}_{\beta} \gamma_{abcd}{}^{\beta}{}_{\alpha} \ ,
\end{aligned}\end{equation}
where $\mu_{ba}=-\mu_{ab}$ and $\ast_8 \mu_{abcd}=\mu_{abcd}$
(the symmetric and anti-self-dual pieces are obtained from the $\rep{70}$ representation $\mu^{\alpha\beta\gamma\delta}$)
and $\mu_{ab}=g_{ac} \mu^c{}_b$.
Similarly, if only the ${\bf 36}$ and the $\bar {\bf 36}$ components are non-zero in the $SU(8)$ decomposition of the ${\bf 912}$, we get the $\SLE$ components ${\bf 36}$ and ${\bf 420}$ in the following way
\begin{equation}\label{36420} \begin{aligned}
\phi^{ab} =& -\tfrac{1}{2} (\phi^{\alpha \beta} + \bar \phi^{\alpha \beta}) C_{\alpha\beta} g^{ab} \ , \\
\phi^{abc}{}_d= & - \tfrac{3}{16}(\phi^{\alpha \beta} + \bar \phi^{\alpha \beta})  (\gamma^{abc}{}_d)_{\alpha \beta} \ , \\
\phi_{ab} =& \tfrac{\iu}{2} (\phi^{\alpha \beta} - \bar \phi^{\alpha \beta}) C_{\alpha\beta} g_{ab} \ , \\
\phi_{abc}{}^d= & \tfrac{3\iu}{16} (\phi^{\alpha \beta} - \bar \phi^{\alpha \beta}) (\gamma_{abc}{}^d)_{\alpha \beta} \ .
\end{aligned}\end{equation}

\section{Technical computations} \label{App:computations}
\subsection{The $N=1$ superpotential}
\label{appsuperpotential}
In this appendix we give the computation of the superpotential given in \eqref{superpotential_SU7}.
We start from \eqref{def_superpotential} where the differential operator is given by \eqref{diff_op}. The form of $\phi$ is given in (\ref{phi_SU7} -- \ref{phi4A6}).
In order to compute \eqref{def_superpotential}, we consider first
\begin{equation} \label{superpotential_DA4}\begin{aligned}
D^C (A^{\ D}_{4 \, B} \phi^A_{DC} +A^{\ D}_{4 \, C} \phi^A_{BD} - A^{\ A}_{4 \, D} \phi^D_{BC}) =&  A^{\ D}_{4 \, B} D^C \phi^A_{DC}- A^{\ A}_{4 \, D} D^C \phi^D_{BC} \\ &
+ (D^C A^{\ D}_{4 \, B} ) \phi^A_{DC} - (D^C A^{\ A}_{4 \, D}) \phi^D_{BC} \\ &
+ (D^C A^{\ D}_{4 \, C}) \phi^A_{BD} +A^{\ D}_{4 \, C} D^C \phi^A_{BD} \ ,
\end{aligned} \end{equation}
where we used $Sp(56)$ indices.\footnote{Note that $E_{7(7)} \subset Sp(56)$.} Now we can translate this back into $E_{7(7)}$ indices, rewriting the above equation as
\begin{equation}
 D \otimesA  (A_4 \cdot \phi) = A_4 \cdot (D \otimesA \phi) + (D \otimes A_4) \times \phi + (D \cdot A_4) \otimesA \phi  +  \big(A_4 \cdot D\big) \otimesA \phi\ ,
\end{equation}
where we used the notation of Eq. (\ref{tensorprodnotation}) and in the last term the differential operator $D$ acts on $\phi$.
If we use the form $A_4 = (0, \rho_8 \wedge A_3)$ and the form of the differential operator \eqref{diff_op}, we find that
\begin{equation}\label{DAzero}
D \cdot A_4  = 0 \ , \qquad A_4 \cdot D = 0 \ .
\end{equation}
Now we can use a variant of the Hadamard formula to find
\begin{equation}\label{Hadamard}
D \phi = \e^{A_4} D (e^{-A_4} \phi) - \e^{A_4} \big(\big[D \otimes A_4\big] \times (e^{-A_4} \phi)\big) + \tfrac12 \e^{A_4} \big( \big(A_4 \cdot \big[D \otimes A_4\big] \big) \times (e^{-A_4} \phi)\big) + \dots \ ,
\end{equation}
where we can compute the coefficients to be
\begin{equation}\label{DAcoeffs} \begin{aligned}
\big[D \otimes A_4\big] = & (0 , 2 (\epsilon \cdot (\rho_8 \wedge \diff A_3)) \otimes (\rho_8),0,0) \ ,\\
A_4\cdot \big[D \otimes A_4\big] = & (0,0 , - 2 \ast_8(A_3 \wedge \rho_8 \wedge \diff A_3) (\rho_8)\otimes(\rho_8) ,0) \ ,\\
A_4\cdot (A_4\cdot \big[D \otimes A_4\big])=& 0 \ .
\end{aligned}\end{equation}
The last equation actually puts all further terms in \eqref{Hadamard} to zero.
From this together with \eqref{def_superpotential}, the superpotential can be computed to be
\begin{equation}
 \e^{K_\phi/2}W = \tfrac12 \iota_{\rho_8} \phi_4 \wedge \diff \phi_4 + \tfrac12\rho_8 \wedge \phi^c_4 \wedge \diff^{\dagger} \phi_4 -2 \iu \phi_4 \wedge \iota_{\rho_8} \diff A_4 + (\iota_{\rho_8} A_4) \wedge \diff A_4 \ .
\end{equation}
In order to make the dependence on $A_6$ explicit, we use Formula \eqref{superpotential_DA4} but now for $\hat A = (\rho_8 \otimes \hat A, 0)$ and note that $D^B \hat A^A{}_B = 0$ and $\hat A^D{}_C D^C = 0$. Therefore, we find
\begin{equation}
D \cdot (\hat A \cdot \phi) = \hat A \cdot (D\otimesA \phi) + [D \otimes \hat A] \times \phi \ .
\end{equation}
Furthermore, we compute
\begin{equation}\label{DA6}
D \otimes \hat A= (0, 0, (\rho_8 \wedge \diff A_6) \rho_8 \otimes \rho_8 , 0 ) \ ,
\end{equation}
giving
\begin{equation}\begin{aligned}
 \e^{K_\phi/2}W = & \tfrac12 \iota_{\rho_8} \phi_4^{(0)} \wedge \diff \phi_4^{(0)} + \tfrac12\rho_8 \wedge \phi_4^{(0)} \wedge \diff^{\dagger} \phi_4^{(0)} +\rho_8 \wedge \diff A_6 \\ & -2 \iu \phi_4^{(0)} \wedge \iota_{\rho_8} \diff A_4  + (\iota_{\rho_8} A_4) \wedge \diff A_4 \ .
\end{aligned}\end{equation}
From self-duality of $\phi_4^{(0)}$ we find that the first two terms actually agree. This means that we finally have
\begin{equation}\label{superpotential}
 \e^{K_\phi/2}W = \iota_{\rho_8} \phi_4^{(0)} \wedge \diff \phi_4^{(0)} + \rho_8 \wedge \diff A_6 -2 \iu \phi_4^{(0)} \wedge \iota_{\rho_8} \diff A_4 + (\iota_{\rho_8} A_4) \wedge \diff A_4 \ .
\end{equation}

\subsection{The $N=2$ prepotentials}
\label{appprepotential}
In this appendix we compute the Killing prepotentials of the $N=2$ theory given in \eqref{prepotentials_sol}. This computation is very similar to the one in Appendix \ref{appsuperpotential}.
We start with the second term in \eqref{prepotentials} where the differential operator is given by \eqref{diff_op}. The objects $K_a$ and $L$ defining the $SU(6)$ structure are given in \eqref{KSU6} and \eqref{LSU6}, respectively.
In order to compute \eqref{prepotentials}, we consider first
\begin{equation}\begin{aligned}
D^B (A^{\ A}_{4 \, D} K^D{}_{B} - A^{\ D}_{4 \, B} K^A{}_{D}) =&  A^{\ A}_{4 \, D}  D^B K^D{}_{B} + (D^B A^{\ A}_{4 \, D}) K^D{}_{B}- A^{\ D}_{4 \, B} D^B K^A{}_{D} \\ & - (D^B A^{\ D}_{4 \, B}) K^A{}_{D} \ ,
\end{aligned}\end{equation}
where we again used $Sp(56)$ indices. Translating this back into $E_{7(7)}$ indices, we find
\begin{equation}
 D (A_4 \cdot K) = A_4 \cdot (D K) + \big[D \otimes A_4 \big] \otimesF K + \big[D \cdot A_4\big] \cdot K  +  \big(A_4 \cdot D\big) \cdot K\ ,
\end{equation}
where  in the last term the differential operator $D$ acts on K. The last two terms vanish due to $A_4 = (0, \rho_8 \wedge A_3)$, cf.\ \eqref{DAzero}.
From the Hadamard formula we find
\begin{equation}\begin{aligned}
D K = & \e^{A_4} D (e^{-A_4} K) - \e^{A_4} \big(\big[D \otimes A_4\big] \otimesF (e^{-A_4} K)\big) \\ & + \tfrac12 \e^{A_4} \big( \big(A_4 \cdot \big[D \otimes A_4\big] \big) \cdot (e^{-A_4} K)\big) \ ,
\end{aligned}\end{equation}
where the coefficients are given in \eqref{DAcoeffs}.
We can treat the transformations $\hat A = (\rho_8 \otimes \hat A, 0)$ using the same equations and \eqref{DA6}, so that
\begin{equation}\begin{aligned}
D K = & \e^{A_{\rm shifts}} D (e^{-A_{\rm shifts}} K) - \e^{A_{\rm shifts}} \big(\big[D \otimes A_4\big] \otimesF (e^{-A_{\rm shifts}} K)\big) \\ & + \tfrac12 \e^{A_{\rm shifts}} \big( \big(A_4 \cdot \big[D \otimes A_4 \big] \big) \cdot (e^{-A_{\rm shifts}} K)\big)\\ & - \e^{A_{\rm shifts}} \big(\big[D \otimes \hat A\big] \otimesF (e^{-A_{\rm shifts}} K)\big)  \ .
\end{aligned}\end{equation}
For the first term in \eqref{prepotentials} we have essentially to determine the scalar derivative $(L, D)$.
From \eqref{LSU6} and \eqref{diff_op}, we get
\begin{equation}\label{vectorderivative}
 (L^{(0)}, D) = - \e^{K_L} \vol_8^{-1} \llcorner (\tfrac{1}{3!} J^3 \wedge \rho_8 \wedge d) = \e^{-K_L} {\cal L}_{\hat v^{(0)}} \ .
\end{equation}
From this together with \eqref{prepotentials}, the Killing prepotentials can be computed to be
\begin{equation}\begin{aligned}
 P^1 =\ & -4 \iu \kappa J \wedge \rho_8 \wedge \diff \Im \Omega^{(0)}_4 - 4 \kappa \iota_v (\Im \Omega^{(0)}_4) \wedge \rho_8 \wedge \diff A_3 \ , \\
 P^2 =\ &\e^{-K_L}(({\cal L}_{\hat v^{(0)}} + \iota_{[\hat v^{(0)},\hat A]})  \Re \Omega^{(0)}_4) \wedge \Im \Omega^{(0)}_4 -\e^{-K_L}(({\cal L}_{\hat v^{(0)}} + \iota_{[\hat v^{(0)},\hat A]}) \Im \Omega^{(0)}_4) \wedge \Re \Omega^{(0)}_4 \\ &
 + 4 \kappa \rho_8 \wedge \diff A_6 - 2 \kappa^2 \e^{K_L} J \wedge J \wedge \rho_8 \wedge v^{(0)} \wedge \diff v^{(0)} \\ &
 + 4 \kappa (A_3 + \iu \kappa v^{(0)} \wedge J) \wedge \rho_8 \wedge \diff A_3 \ , \\
 P^3 =\ & -4 \iu \kappa J \wedge \rho_8 \wedge \diff \Re \Omega^{(0)}_4 - 4 \kappa \iota_v (\Re \Omega^{(0)}_4) \wedge \rho_8 \wedge \diff A_3 \ .
\end{aligned} \end{equation}
where we also used $[{\cal L}_{\hat v^{(0)}}, \iota_{\hat A}] = \iota_{[\hat v^{(0)},\hat A]}$.



\end{document}